\documentclass[letterpaper,twocolumn,10pt]{article}
\usepackage{zhanggroup}

\usepackage{hyperref}
\hypersetup{
  colorlinks,
  linkcolor={blue!70!green},
  citecolor={green!70!blue},
  urlcolor={orange!70!red}
}
\usepackage{xspace}
\usepackage{cite}
\usepackage{url}

\usepackage{amsmath,amssymb,amsfonts}
\usepackage{algorithmic}
\usepackage{graphicx}
\usepackage{textcomp}
\usepackage{xcolor}
\usepackage{tikz}
\usepackage{array}
\usepackage{pifont}
\usepackage[normalem]{ulem}
\usepackage{multirow}
\usepackage{subcaption}
\usepackage{caption}
\usepackage{booktabs}
\usepackage{makecell}
\usepackage{tcolorbox}
\usepackage[absolute]{textpos}

\graphicspath{ {images/} }
\usepackage{enumitem}
\setlist[itemize]{leftmargin=*}
\usepackage{titlesec}

\newcommand{\mypara}[1]{\smallskip\noindent\textbf{#1.}\xspace}
\newcommand{\database}{\textsc{SecurityNet}\xspace}
\begin{document}

\begin{textblock}{13}(1.5,1)
\centering
To Appear in the 33rd USENIX Security Symposium, August 2024.
\end{textblock}

\title{\database: Assessing Machine Learning Vulnerabilities on Public Models}

\date{}

\author{
{\rm Boyang Zhang}\ \ \
{\rm Zheng Li}\ \ \
{\rm Ziqing Yang}\ \ \
{\rm Xinlei He}\ \ \
{\rm Michael Backes}\ \ \
{\rm Mario Fritz}\ \ \
{\rm Yang Zhang}\ \ \
\\
\\
\textit{CISPA Helmholtz Center for Information Security}
}

\maketitle

\begin{abstract}

While advanced machine learning (ML) models are deployed in numerous real-world applications, previous works demonstrate these models have security and privacy vulnerabilities.
Various empirical research has been done in this field.
However, most of the experiments are performed on target ML models trained by the security researchers themselves.
Due to the high computational resource requirement for training advanced models with complex architectures, researchers generally choose to train a few target models using relatively simple architectures on typical experiment datasets.
We argue that to understand ML models' vulnerabilities comprehensively, experiments should be performed on a large set of models trained with various purposes (not just the purpose of evaluating ML attacks and defenses).
To this end, we propose using publicly available models with weights from the Internet (public models) for evaluating attacks and defenses on ML models.
We establish a database, namely \database, containing 910 annotated image classification models.
We then analyze the effectiveness of several representative attacks/defenses, including model stealing attacks, membership inference attacks, and backdoor detection on these public models.
Our evaluation empirically shows the performance of these attacks/defenses can vary significantly on public models compared to self-trained models.
We share \database with the research community\footnote{We publish \database at \url{https://github.com/SecurityNet-Research/SecurityNet}.} and advocate researchers to perform experiments on public models to better demonstrate their proposed methods' effectiveness in the future.
\end{abstract}

\section{Introduction}
\label{section:introduction}

Machine learning (ML) has been gaining momentum in multiple fields and achieving success in real-world deployments.
However, in recent years, researchers have shown that ML models are vulnerable to various security and privacy risks, such as membership inference~\cite{SSSS17}, model stealing/extraction~\cite{TZJRR16}, and backdoor~\cite{GDG17}.
Quantifying and mitigating ML models' vulnerabilities thus become increasingly important topics.

Currently, most of the research in this field focuses on proposing different attacks and countermeasures.
To evaluate these methods, the common practice is that the researchers train models by themselves and treat these models as potential victims' models (target models) in the experiments.
Based on some of the well-known papers on two popular attacks against ML models, including membership inference and model stealing~\cite{NSH18,JSBZG19,CYZF20,ZRWRCHZ21,LZ21,PTC18,SZHBFB19,SSM19,LWHSZBCFZ22,WG18,HXLHX21}, we find that all of them conduct experiments on target models trained from scratch by the authors.

This practice, however, faces several limitations.
The behavior of the models can vary greatly on different architectures and different datasets.
Since training state-of-the-art ML models is resource-intensive and time-consuming, the target models used in machine learning security and privacy research tend to be limited to the most popular architectures trained on the common and approachable experiment datasets (e.g., CIFAR-10~\cite{CIFAR}, CIFAR-100~\cite{CIFAR}, and SVHN~\cite{SVHN}).
Also, the number of models used in the evaluation is often small.

Furthermore, even with the same model architecture and dataset, different procedures and hyperparameters used for training can still drastically alter the model's behavior.
For state-of-the-art models, huge efforts are dedicated to fine-tuning hyperparameters to find the best training procedures, thus maximizing the chosen model architecture's potential on the target task.
Since research in security and privacy tends to focus on developing attacks and countermeasures, it is unrealistic for the researchers to have a similar level of dedication to training their victim models.
Subsequently, the victim model in the experiments might not be adequately trained, whereby the model's performance on the target task is lower than the given architecture's best result (we show evidence in \autoref{section:database}).

Publishing models with weights on the Internet (\emph{public models}) is becoming a common practice in the machine learning community to increase research reproducibility and provide benchmarks on different ML tasks.
These public models cover a wide variety of model architectures and datasets.
Moreover, many of these public models are already integrated into companies' products deployed in the real world.
For instance, certain transformer models on Hugging Face\footnote{\url{https://huggingface.co/}.} have been integrated into Amazon SageMaker.\footnote{\url{https://aws.amazon.com/machine-learning/hugging-face/}.}
\emph{To fully assess the effectiveness of different attacks and defenses on machine learning models, we argue that the experiments should be conducted on such public models when possible.}

\subsection{Our Contributions}

In this work, we take the first step towards conducting ML models' security and privacy vulnerability evaluation on public models.
We collect a large-scale dataset of public models, namely \database, to evaluate three popular attacks/defenses in this field, including membership inference attack, model stealing attack, and backdoor detection.
We omit the popular topic of evasion attacks due to existing benchmarks~\cite{CASDFCMH21,HTCH22}, but we do include baseline results and discussion in \autoref{section:evasion_attacks}.
Note that we focus on image classification models as they have been extensively studied by the trustworthy machine learning research community.

\mypara{\database}
We build a public model database \database by collecting a large number of public models used for image classification from multiple open-source platforms on the Internet, such as Paper with Code~\cite{PapersWithCode}, Kaggle~\cite{Kaggle}, and GitHub~\cite{GitHub}.
Many of our public models come from machine learning libraries that contain models with various architectures trained on multiple datasets.
These models are usually trained for performance benchmarks, so they have as high as possible prediction accuracy on the target task (of the given architectures).
We refer to such models as \emph{benchmark models}.
We further manually search for publicly available models from research papers published in top-tier security, machine learning, and computer vision conferences.
Among these models from the research papers, we refer to those related to the topic of machine learning security and privacy as \emph{security models}, and the rest are considered as part of the benchmark models.
We notice that most of the security models are used for adversarial example research.
In general, benchmark and security models will be the main focus of our analysis.

\database comprises 910 models covering 42 different datasets.
For each model in \database, we further annotate its relevant information from three dimensions, namely dataset (e.g., sample size, split ratio, topic category, class fidelity, etc.), model (e.g., number of parameters, FLOPs, architecture type, etc.), and metadata (e.g., publisher type, published year, venue, model purpose, etc.).
We hope this information can help security researchers find appropriate public models promptly.
Also, we will continue enlarging \database with newer models to monitor whether ML models are more or less susceptible to the attacks investigated over time.
We plan to share \database with the research community to facilitate the research in machine learning security and privacy.

\mypara{Evaluation Results}
With the help of \database, we are able to perform an extensive analysis of model stealing, membership inference, and backdoor detection on a large set of public models.
To the best of our knowledge, this has not been done before.
Our experiments confirm some known results from the literature (but on a much larger number of models), uncover some new insights, and show that attacks and defenses can behave differently on public models than on researchers' self-trained models.

For model stealing on benchmark models from \database, we find using a larger and more complex surrogate model architecture to limit the difference between the surrogate and victim models does not improve the attack performance, which differs from previous results~\cite{OSF19}.
Also, we observe a negative correlation between the attack performance and the victim model's target task performance.
Such a negative correlation has been observed in~\cite{JCBKP20} previously; our experiments on a much larger number of (public) models further confirm this.
In addition, if the target model is too complex (we test model stealing on a RegNetY-320 model~\cite{RKGHD20} with 145 million parameters for the first time), model stealing is ineffective.
The public models trained for security and privacy research (security models) typically perform much worse on their target tasks than the benchmark models.
Interestingly, for the low-performing security models, we observe a positive correlation between the attack and the target task performance (the opposite of our observation on benchmark models).
All these new insights demonstrate the benefits of performing model stealing experiments on public models.

For membership inference, we evaluate two types of attack methods, namely metric-based attacks~\cite{SM21} and MLP-based attacks~\cite{SSSS17,SZHBFB19,NSH19}.
We empirically confirm some results shown in previous works~\cite{SSSS17}, e.g., the attack performance positively correlates with the model's overfitting level on the target task on the large-scale public models.
On the other hand, we also discover that the attack methods can be dataset-dependent.
For instance, previous works show that the MLP-based attack using full posteriors as its input has the same performance as using the top-k (e.g., top-3) posteriors~\cite{SZHBFB19}.
However, on datasets with a large number of classes, e.g., 1,000 classes (ImageNet-1k~\cite{DDSLLF09}), we show that using top-3 posteriors as inputs achieve much better performance than full posteriors (on average, the attack AUC increases by 5.1\%).
Some of the security models with lower performance on their target tasks appear to be less vulnerable to membership inference attacks than benchmark models, even when they share a similar overfitting level.
We do not make the same observation on security models that achieve similar target task performance as benchmark models.

Finally, we examine three backdoor detection techniques on the public models from \database: Neural Cleanse~\cite{WYSLVZZ19}, Strong Intentional Perturbation (STRIP)~\cite{GXWCRN19}, and NEO~\cite{UPWLRC22}.
Assuming all the benchmark models collected are non-backdoored, we discover that Neural Cleanse has high false positive rates on benchmark models trained on CIFAR-10 and SVHN (20.9\% and 13.7\%, respectively).
By manually checking the generated trigger images, we confirm that the detected triggers are indeed falsely identified.
On the other hand, both STRIP and NEO are more robust.
They successfully avoid labeling any clean inputs as backdoored samples on CIFAR-10 and SVHN models in our experiments.
This again shows the necessity of evaluating attacks and defenses on public models.

\mypara{Implications}
This work aims to provide a more realistic overview of the landscape of machine learning attacks and defenses.
We also want to point out that some of the current evaluation results on researchers' self-trained models might not generalize to public models.
We advise researchers to examine their proposed methods on at least a few public models for more comprehensive evaluations in the future.
Hence, we will share \database.
We hope our annotations and experiment results on baseline attacks/defenses will greatly minimize the effort for researchers to find appropriate public models for their purpose.

\section{\database}
\label{section:database}

One of the main contributions of our work is \database, a database containing publicly available models with weights.
We focus on one of the most popular machine learning tasks, image classification, as it is also typically used to demonstrate the effectiveness of attacks and defenses on ML models.

\begin{figure*}[!t]
\centering
\begin{subfigure}{0.33\textwidth}
\centering
\includegraphics[width=\textwidth]{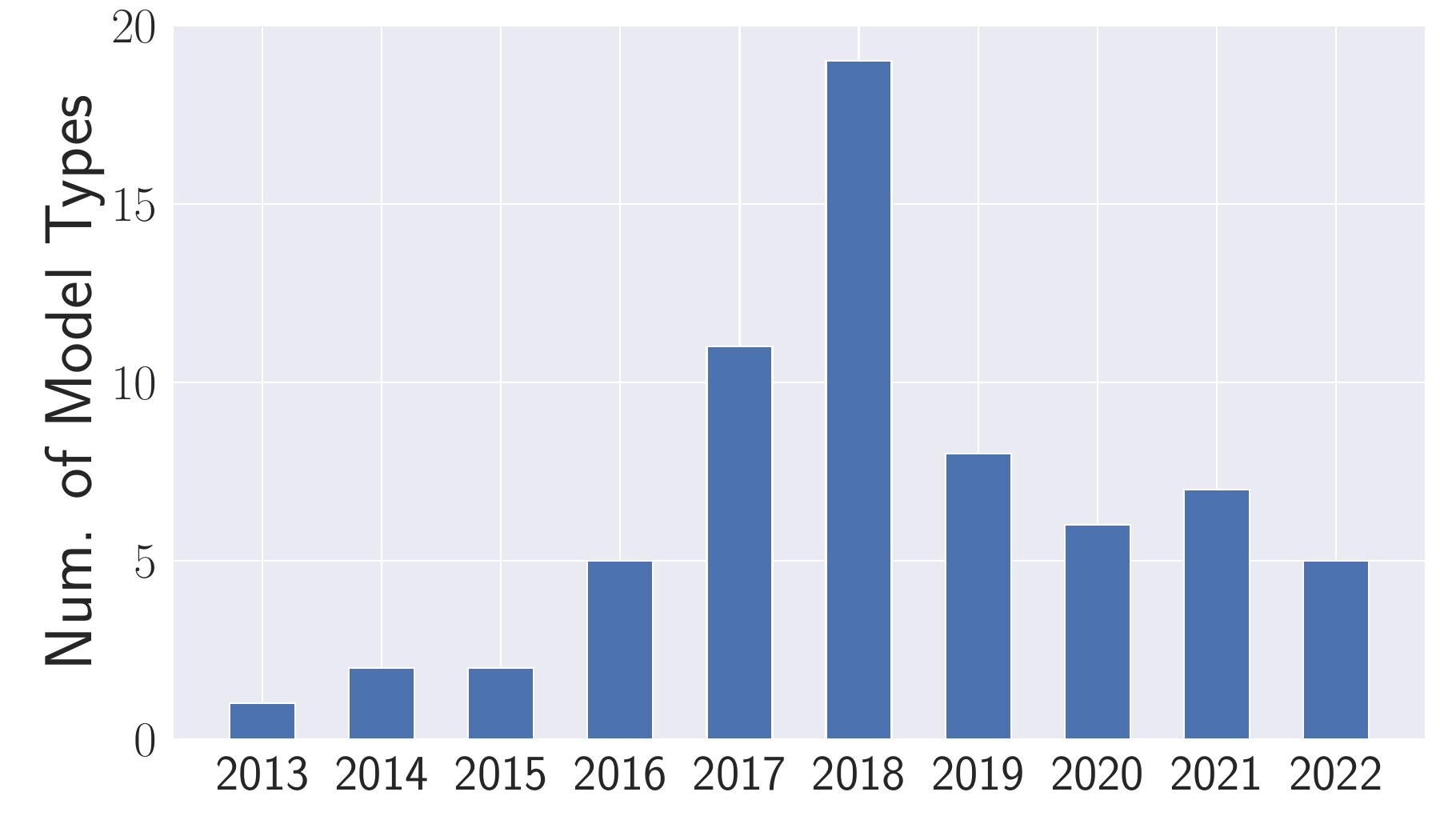}
\caption{Publishing Year of Benchmark Models}
\label{figure:database_year}
\end{subfigure}%
\hfill
\begin{subfigure}{0.33\textwidth}
\centering
\includegraphics[width=\textwidth]{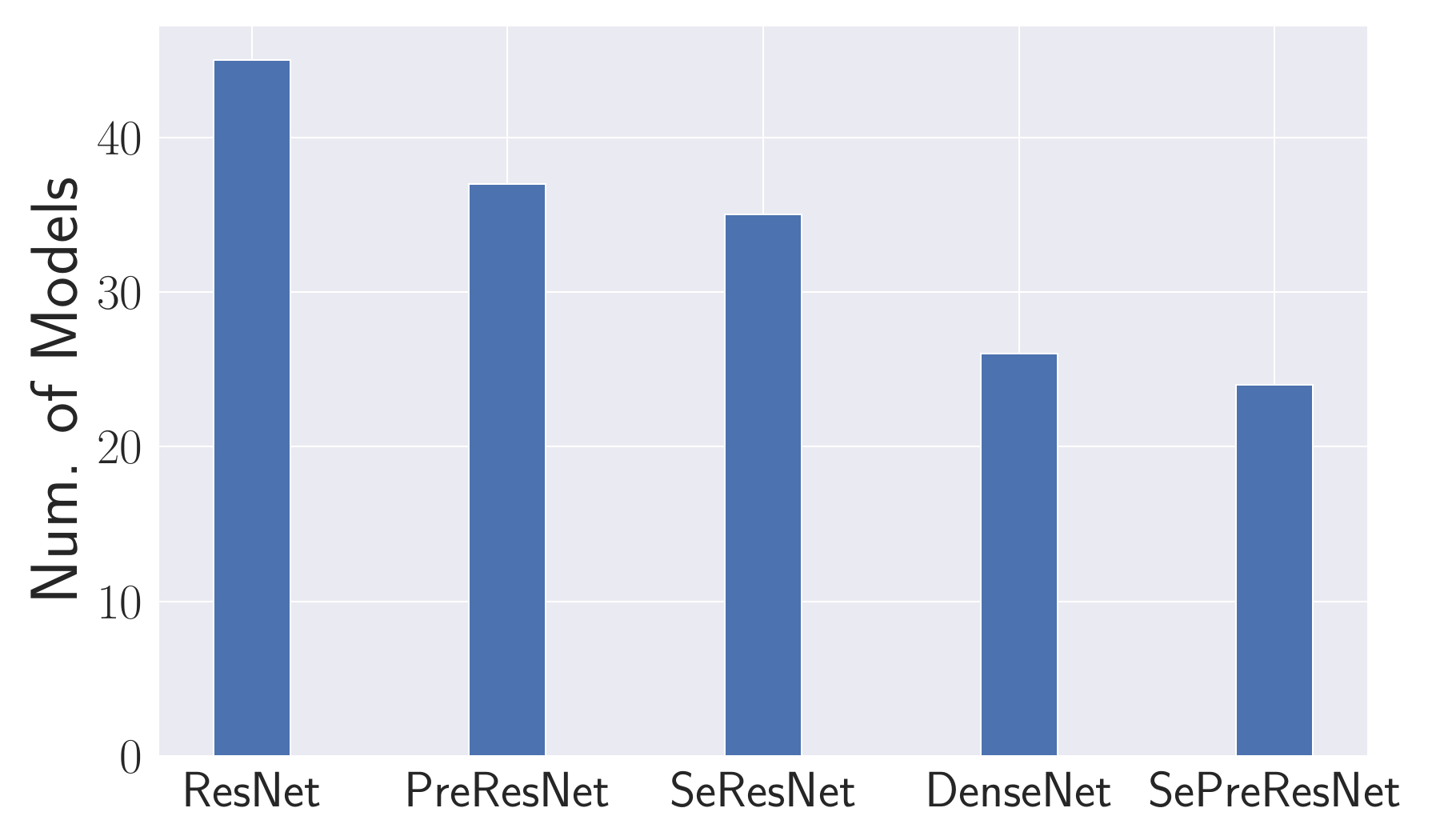}
\caption{Top Benchmark Model Types}
\label{figure:type_benchmark}
\end{subfigure}%
\hfill
\begin{subfigure}{0.33\textwidth}
\centering
\includegraphics[width=\textwidth]{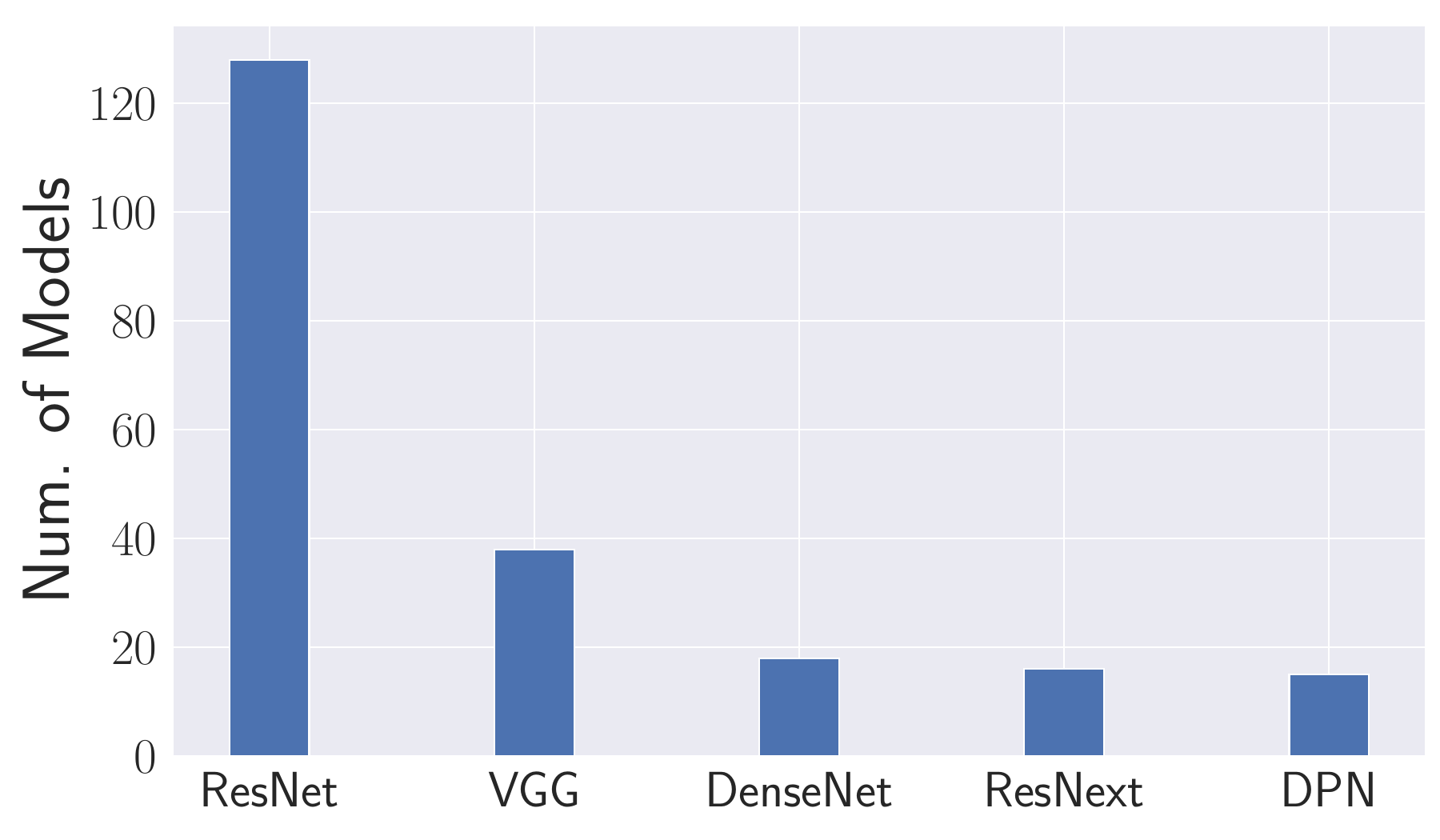}
\caption{Top Security Model Types}
\label{figure:type_security}
\end{subfigure}%
\hfill
\begin{subfigure}{0.33\textwidth}
\centering
\includegraphics[width=\textwidth]{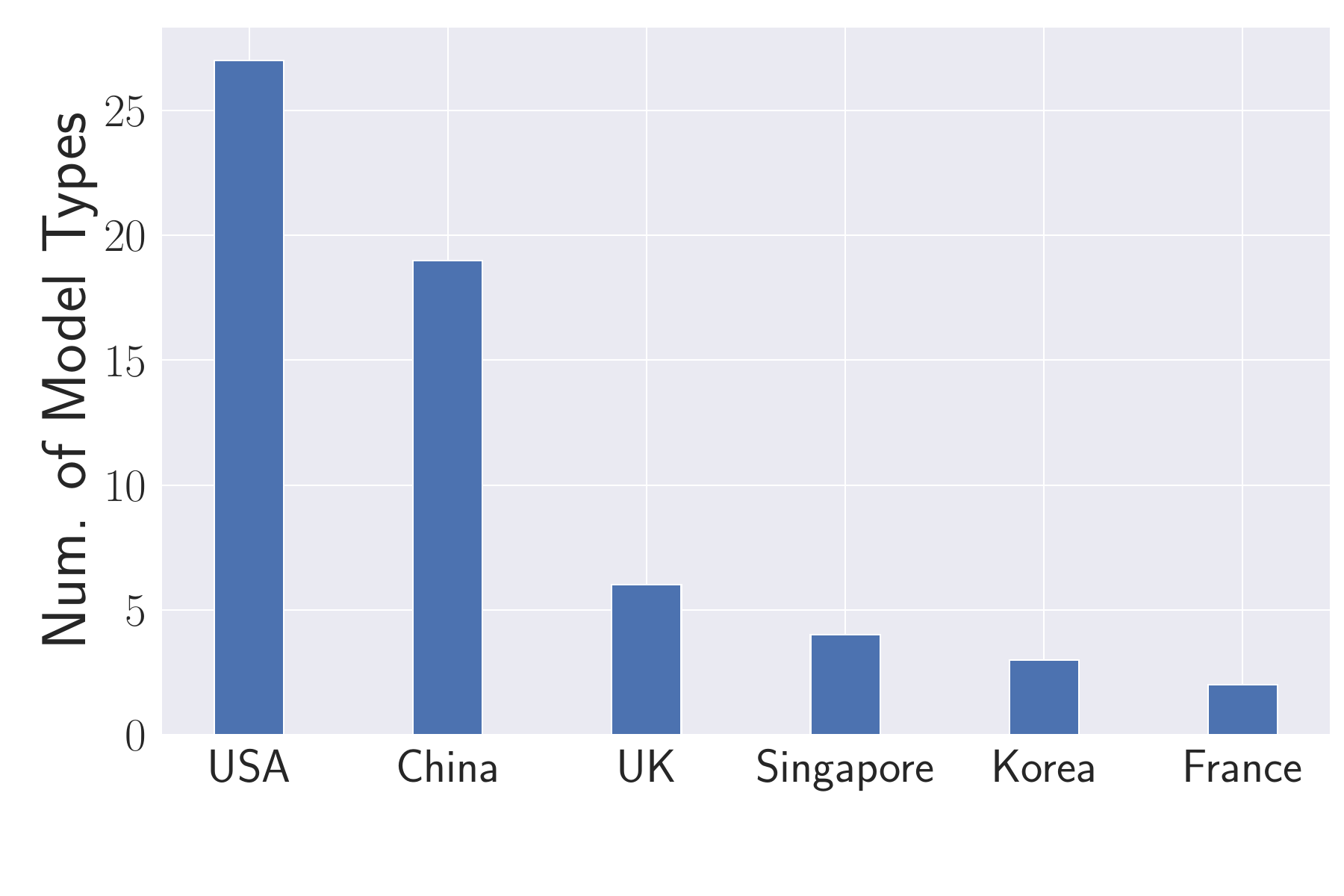}
\caption{Top Countries Regarding Model Type}
\label{figure:database_country}
\end{subfigure}%
\hfill
\begin{subfigure}{0.33\textwidth}
\centering
\includegraphics[width=\textwidth]{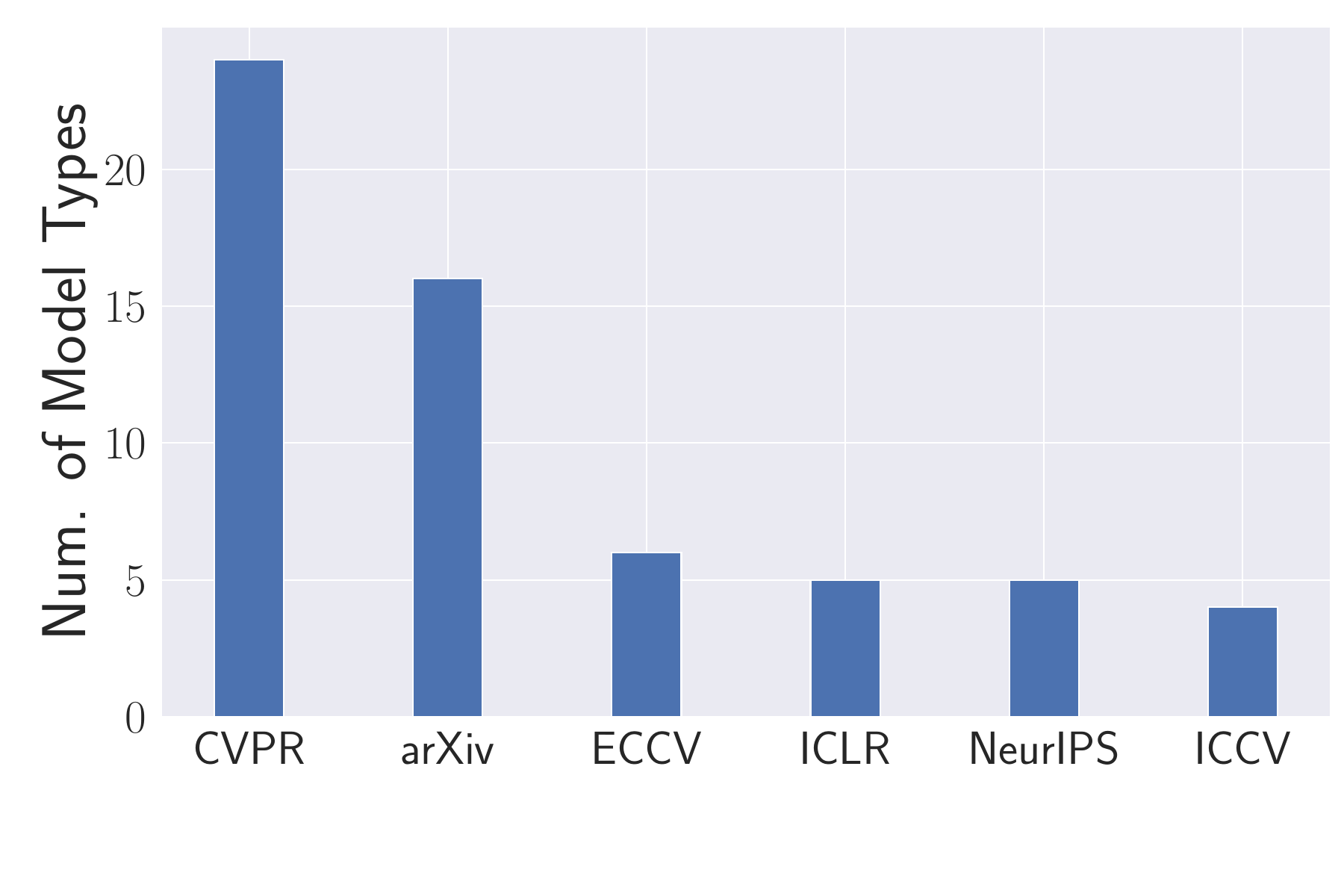}
\caption{Top Venues Regarding Model Type}
\label{figure:database_conference}
\end{subfigure}%
\hfill
\begin{subfigure}{0.33\textwidth}
\centering
\includegraphics[width=\textwidth]{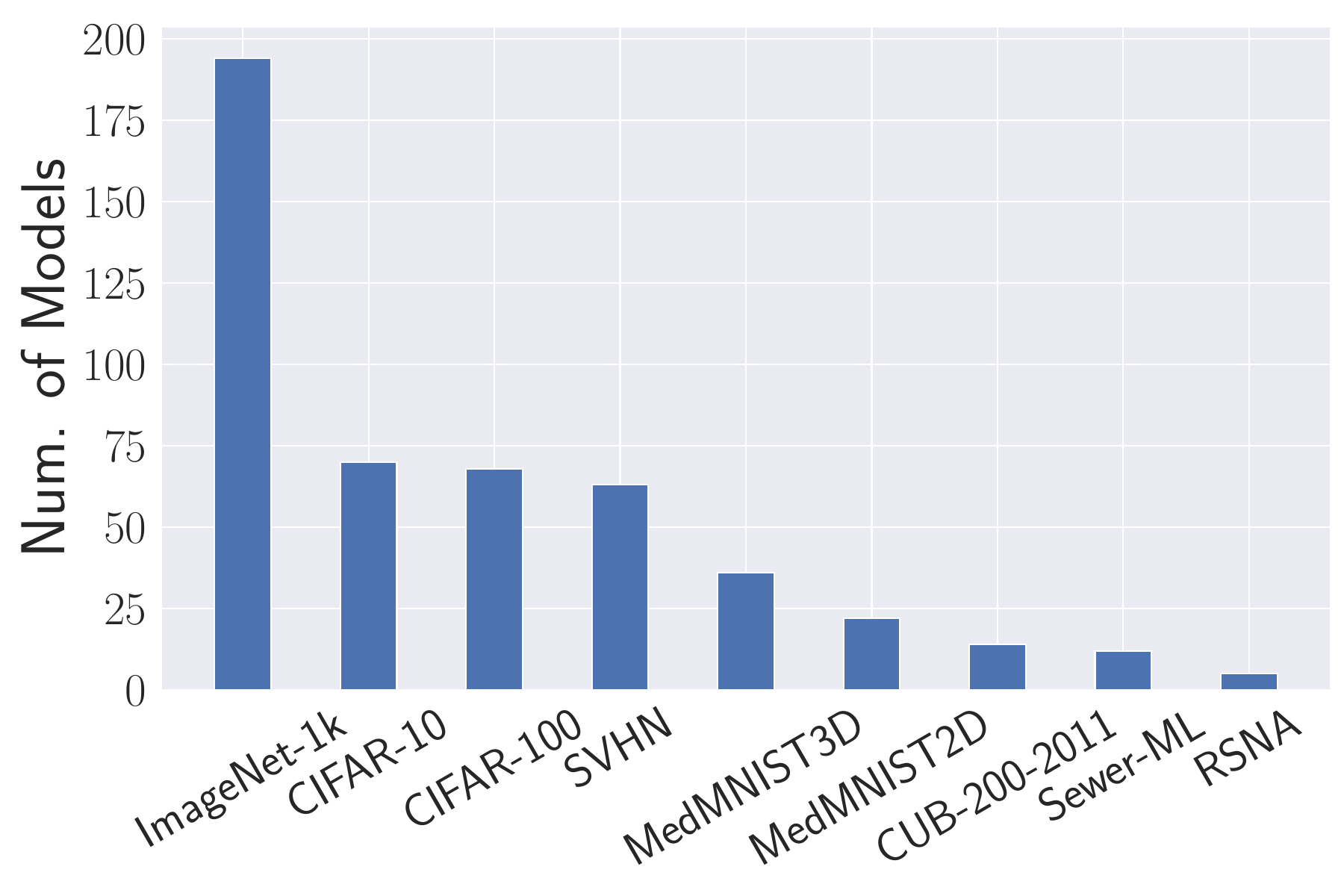}
\caption{Top Benchmark Models' Datasets}
\label{figure:database_dataname}
\end{subfigure}%
\caption{\database statistics.}
\label{figure:database}
\end{figure*}

\subsection{Model Collection}
\label{subsection:model collection}

\mypara{Datasets to Models}
Our main model collection process, namely datasets to models, consists of two steps, namely dataset searching and model collection.
In a nutshell, we first find a diverse set of datasets and then collect public models trained on these datasets.

For dataset searching, we focus on image datasets that are mainly used for classification tasks.
Note that we also plan to extend to other types of tasks in the future.
The diversity of the datasets is critical.
We consider two sources for dataset collection, namely Paper with Code~\cite{PapersWithCode} and Kaggle~\cite{Kaggle}.
Paper with Code is a website that provides open-source content, including machine learning papers, codes, datasets, methods, and evaluation results.
The website's collection of datasets covers the majority of datasets commonly used for machine learning research.
Kaggle is a crowd-sourcing platform that is popular in the data science community.
It is well-known for hosting data science competitions and challenges in cooperation with many companies and research institutes.
The datasets used in these competitions have a huge variety and typically differ from the experiment datasets collected from Paper with Code.

After collecting a variety of datasets, we then use these datasets as a starting point to search for publicly available models trained on these datasets.
Our search can be summarized in a few directions.
First and foremost, pre-trained model libraries are some of the most valuable sources, such as PyTorch's official torchvision library.\footnote{\url{https://github.com/pytorch/vision}.}
They typically contain a wide range of popular models that are trained to have high performance on the target tasks.
We refer to these models as \emph{benchmark models}.
Models from reputable sources also provide high confidence in their qualities.
These qualities, including no additional data used, a proper partition of training and test data, no malicious data (e.g., backdoor triggers), etc., are especially important for our later analysis.
These model libraries, however, also suffer some downsides.
Specifically, these libraries typically contain only well-established architectures and benchmark experiment datasets.

Furthermore, we extend our benchmark model collection to other sources, such as models from Kaggle competitions and Paper with Code.
These sources provide a much wider range of models in several dimensions, such as model variety, purpose, and quality.
We emphasize here that due to the wide range of sources in our model collection, the quality of the models cannot be guaranteed in the same way as with reputable sources (e.g., the Pytorch torchvision library).
However, this variation in model quality yields a valuable comparative dimension to our analysis.\footnote{For simplicity, these models are also considered as benchmark models in \database.}

Additionally, there are models that we do not include in the database.
We exclude models with corrupted weights that cannot be loaded properly or have extremely low performance.
Moreover, the same model can appear on multiple platforms.
We also exclude these duplicates to avoid over-representation of the same model.

\mypara{Academic Papers to Models}
Furthermore, to obtain an overview of the current models used in research, we manually search for image classification models provided by authors of papers published in recent top-tier conferences.
We consider the following security \& privacy, machine learning, and computer vision conferences in the last four years: IEEE S\&P, USENIX Security, ACM CCS, NDSS, NeurIPS, ICML, ICLR, CVPR, ICCV, and ECCV.
To simplify the process, for machine learning and computer vision conferences, we directly search on GitHub for the corresponding repositories (e.g., using keywords ``CVPR 2022'').
Considering the popularity of GitHub, we believe we capture the majority of the published models in those venues.
Meanwhile, for security conferences, we manually check all papers to obtain the models.

We also especially focus on models from papers on the topic of trustworthy machine learning from all the conferences considered and refer to them as \emph{security models}.
During our collection process, we noticed that the majority of the security models are derived from papers on adversarial example studies.
Unfortunately, we cannot find any public models on model stealing and membership inference.\footnote{Note that for backdoor attacks, our goal is to apply the current backdoor detection methods to the public models.
Thus, we do not include the backdoored models published with backdoor-related research papers in our database.}
Overall, the addition of these models will allow us to conduct a more comprehensive analysis in \autoref{section:evaluation results}.
Note that the models from research papers that are not related to security and privacy are considered benchmark models as well.

\subsection{Annotation}
\label{subsec:annotation}

Once having collected the models, we then annotate their relevant information.\footnote{We explain our annotation process in detail in \autoref{section:categories}.}
The annotation serves two purposes.
First, it provides a guideline for our analysis of the current landscape of security and privacy attacks/defenses.
Second, it serves as a valuable source for future research.

The annotation for models in \database includes quantitative information about their training sets, such as the number of classes, size, image dimensions, etc.
Furthermore, we record qualitative information, such as topic categories.
Our models' datasets cover a broad range of topics, including natural scenery, medical scans, traffic signs, satellite images, and more.
Within each topic category, we further distinguish the dataset's class granularity.
For example, while ImageNet-1k and CUB-200-2011 are both categorized as natural scenery, ImageNet-1k includes multiple types of objects ranging from different animals to cars and park benches.
We label datasets like ImageNet-1k as coarse-grained datasets.
CUB-200-2011 only contains images of different types of birds.
Therefore, it is labeled as a fine-grained dataset.

Besides models' datasets, we further annotate their intrinsic properties and metadata.
Intrinsic model properties include the number of parameters, FLOPs, architecture type, presence of certain elements (e.g., dropout, batch normalization), etc.
We also annotate the models' metadata, such as publishing venues (for research-based models), number of authors, etc.
Given the large number of models we have collected, we can analyze different attacks and defenses from the metadata dimension, which, to our knowledge, has not been done before.
\autoref{section:categories} in the appendix provides the complete list of categories annotated by us for \database.

\subsection{Summary}

In total, \database contains 910 public models, of which 665 are benchmark models and 245 are security models.
These models are trained over 42 different datasets from 13 categories.
The models cover an extensive set of 220 architectures (e.g., ResNet-18~\cite{HZRS16}, ResNet-50, DLA-169~\cite{YWSD18}, BagNet-33~\cite{BB19}, etc.) based on 60 different model types (ResNet, DLA, BagNet, etc.).
The oldest model type included was first introduced in 2012~\cite{KSH12} and the latest one~\cite{LMWFDX22} in 2022.
Note that for benchmark models, we record the year when the model type was first introduced instead of the trained models' publishing time.
\autoref{figure:database} shows some general statistics of the models in \database.

Model collection for \database will be a continuous process.
We plan to update \database on a bi-annual basis to add new publicly available ML models.
This allows us to keep tracking ML models' security and privacy vulnerabilities over time.
We will also make \database easily accessible to the research community.

\begin{figure}[!t]
\centering
\begin{subfigure}{0.15\textwidth}
\centering
\includegraphics[width=\textwidth]{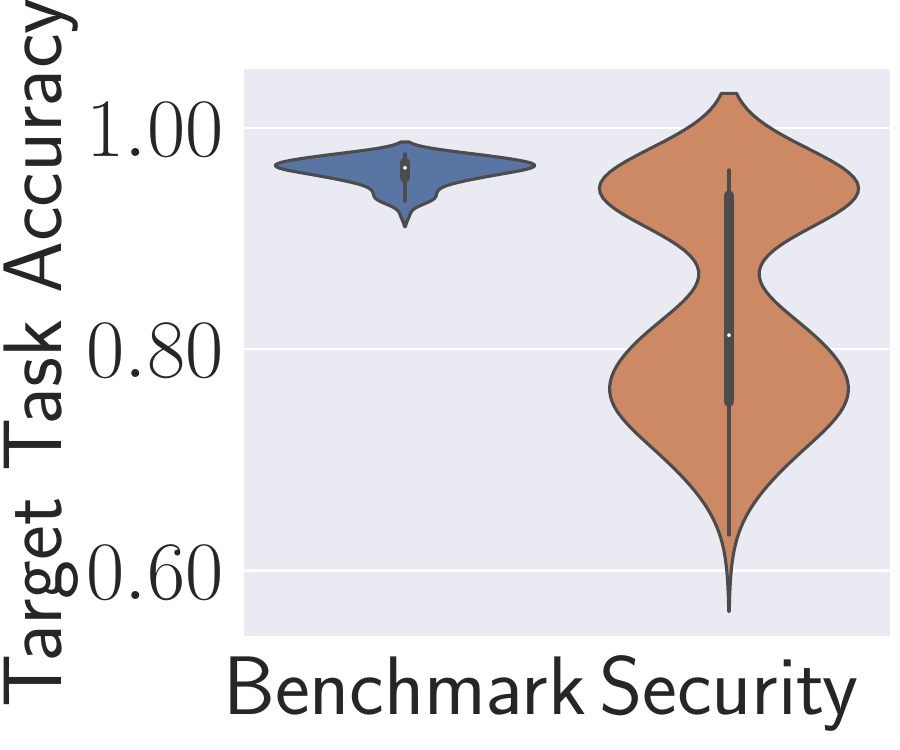}
\caption{CIFAR-10}
\label{figure:cifar10_target_accuracy}
\end{subfigure}%
\hfill
\begin{subfigure}{0.15\textwidth}
\centering
\includegraphics[width=\textwidth]{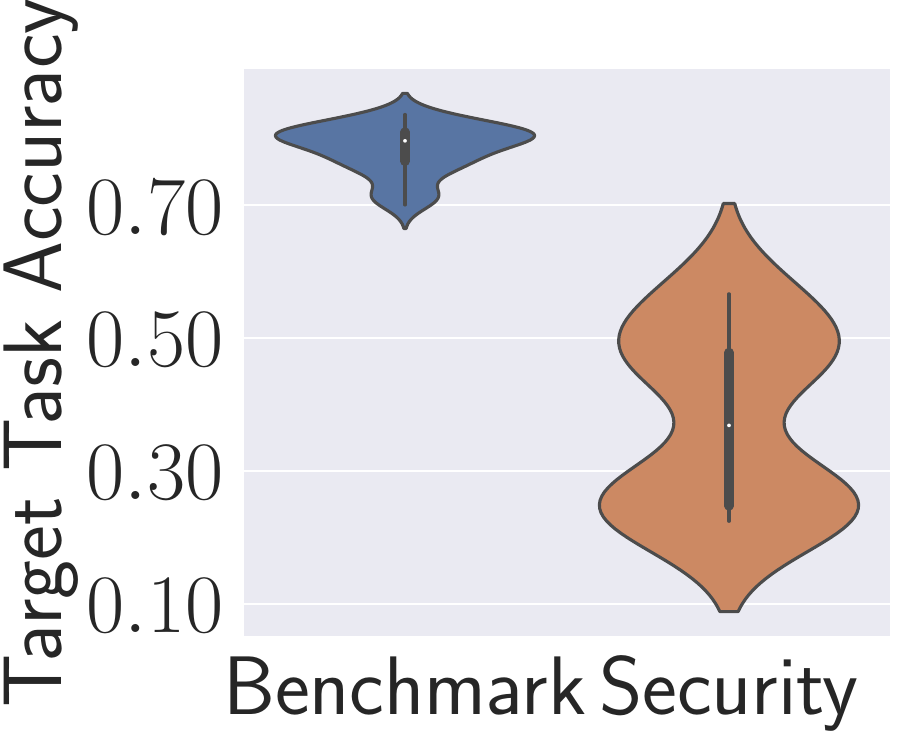}
\caption{CIFAR-100}
\label{figure:cifar100_target_accuracy}
\end{subfigure}%
\hfill
\begin{subfigure}{0.15\textwidth}
\centering
\includegraphics[width=\textwidth]{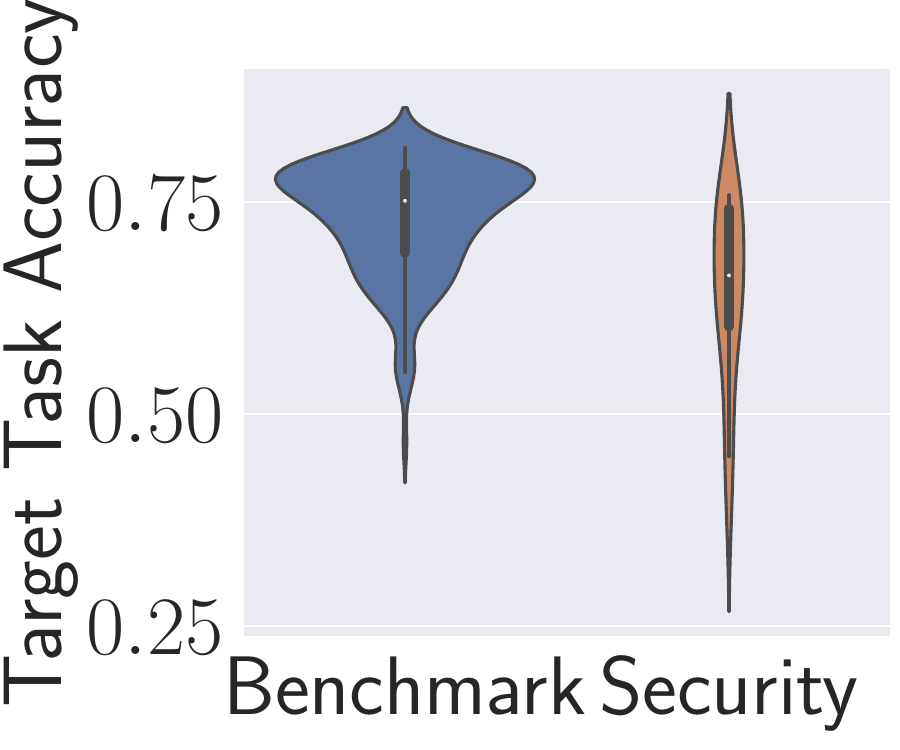}
\caption{ImageNet-1k}
\label{figure:imagenet_target_accuracy}
\end{subfigure}%
\caption{The model's target task performance with respect to benchmark models and security models.}
\label{figure:violin_target_accuracy}
\end{figure}

\begin{figure}[!t]
\centering
\begin{subfigure}{0.15\textwidth}
\centering
\includegraphics[width=\textwidth]{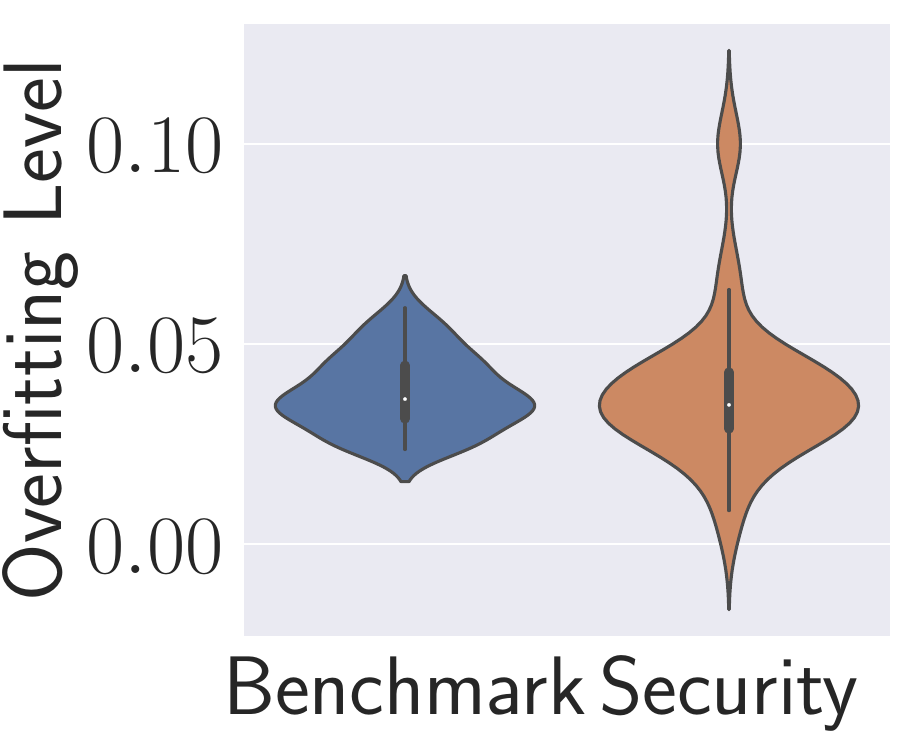}
\caption{CIFAR-10}
\label{figure:cifar10_overfit}
\end{subfigure}%
\hfill
\begin{subfigure}{0.15\textwidth}
\centering
\includegraphics[width=\textwidth]{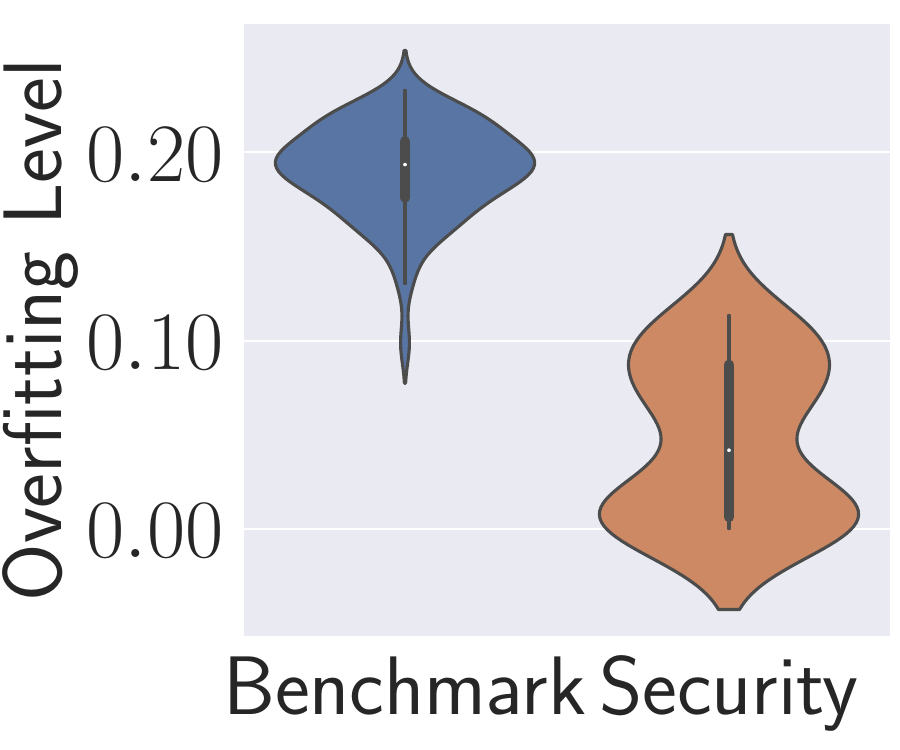}
\caption{CIFAR-100}
\label{figure:cifar100_overfit}
\end{subfigure}%
\hfill
\begin{subfigure}{0.15\textwidth}
\centering
\includegraphics[width=\textwidth]{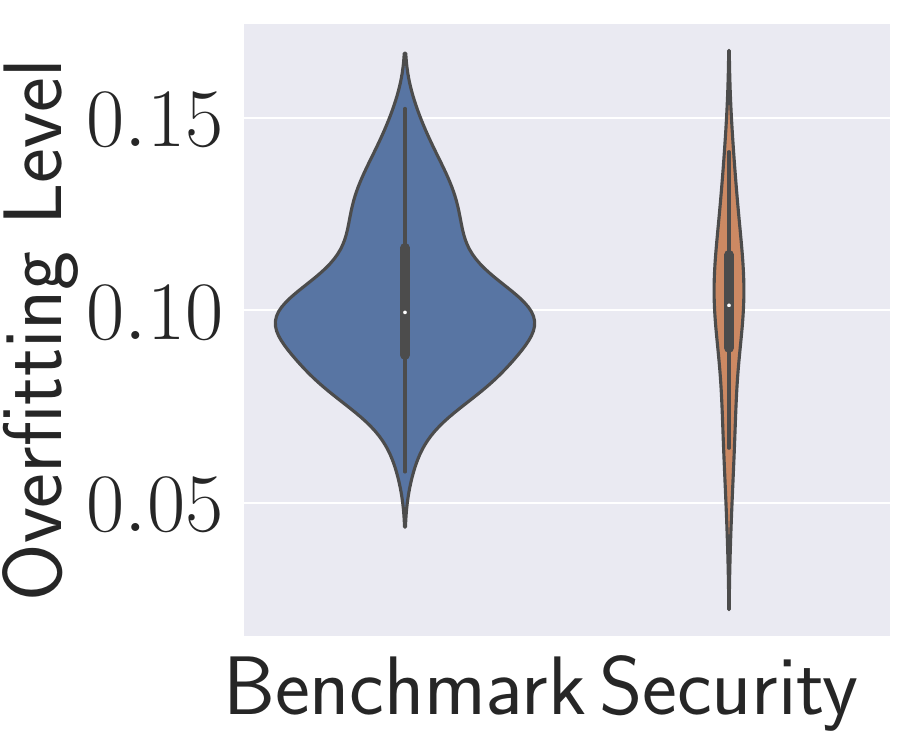}
\caption{ImageNet-1k}
\label{figure:imgnet_overfit}
\end{subfigure}%
\caption{The model's overfitting level with respect to benchmark models and security models.}
\label{figure:overfitting_level}
\end{figure}

\mypara{Security Models vs.\ Benchmark Models}
Based on the models collected, we first observe that the majority of the security models are trained on small experiment datasets, such as CIFAR-10, CIFAR-100, and SVHN.
Only a small amount of the papers include results on more complex datasets like ImageNet-1k.
These four datasets also cover the majority of the papers on security and privacy research that we have found.
Besides, the model architectures used in security models are also limited, e.g., the majority of the architectures are the simpler and popular ones, such as ResNet-18, VGG-16, etc.
\autoref{figure:violin_target_accuracy} shows the model performance difference between the benchmark models and the security models trained on CIFAR-10, CIFAR-100, and ImageNet-1k.
We only use benchmark models that share similar architectures with the security models so that the inherent differences in model architectures do not affect the comparison.
For CIFAR-10 and CIFAR-100 (too few models for ImageNet-1k), we notice that there are two distinct clusters of models, and both have lower performance than benchmark models.
From \autoref{figure:overfitting_level}, we can observe the overfitting gaps between benchmark models and security models are not too different for CIFAR-10 and ImageNet-1k models, while CIFAR-100 security models have a lower overfitting level in general.
Note that the security models here are for adversarial example research, and we cannot find any published target models for model stealing and membership inference.
Nevertheless, some of the membership inference and model stealing papers' reported target task accuracy is still lower than the benchmark models' accuracy in \database.
For instance, CIFAR-100 models from the popular papers in \autoref{section:introduction} have an average accuracy of 69.0\%, compared to our benchmark models' average accuracy of 78.5\%.
Additionally, we find the performance gap still exists in recently published papers (CIFAR-10: 79\%~\cite{CSSWZ22}; CIFAR-10: 77\%, CIFAR-100: 20\%~\cite{YMMBS22}).
We believe these security models are not trained to the architecture's maximum ``potential'' due to limited hyperparameter tuning efforts.
For simplicity, previous works~\cite{LWHSZBCFZ22,AF20} in the security/privacy domain only use one set of batch size, learning rate, optimizer, etc., without further hyperparameter fine-tuning.
In conclusion, the above results demonstrate that some security models are not adequately trained compared to benchmark models.

\section{Attack Methodology and Evaluation Setup}
\label{section:methods}

In this paper, we study three types of attacks/defenses on machine learning models, namely model stealing attack, membership inference attack, and backdoor detection.
They are among the most well-explored subjects in the field of machine learning security and privacy.
In the future, we plan to extend our analyses to other types of attacks/defenses with models from \database.

\subsection{Model Stealing}

\mypara{Threat Model} 
In this attack, the adversary aims to build a surrogate model that mimics the target model's behavior.
Following one of the most popular attacks~\cite{TZJRR16}, we consider an adversary with black-box access to the target model that outputs the full posterior.
The adversary also has access to an auxiliary dataset for querying the target model.
Note that the auxiliary dataset does not necessarily come from the same distribution as the original training set.

\mypara{Methodology}
The adversary first initiates a surrogate model which can adopt a different architecture than the target model~\cite{OSF19}.
Then, the adversary queries the samples from their auxiliary dataset to the target model and gets the output posteriors.
In the end, the adversary trains their surrogate model leveraging the posteriors as ground truth.

In our experiments, we adopt ResNet-18 to initiate the surrogate models.
We further show in \autoref{subsec:eval_model_stealing} that a more complex surrogate model does not increase the attack performance.
We consider two settings for the auxiliary datasets, i.e., partial training set\footnote{We use 50\% of the total training data.
We evaluate how different query budget affects the attack performance in \autoref{section:steal_budget}.} from the target model's training set (default setting) and a large out-of-distribution dataset (specifically, a subset of ImageNet-1k).
We train ResNet-18 for 30 epochs using an SGD optimizer with a learning rate of 0.01.

\mypara{Metrics}
We adopt two most widely-used metrics, namely attack accuracy and attack agreement~\cite{JCBKP20}, in the evaluation.
Accuracy measures the performance of the surrogate model on the original task, while agreement calculates the prediction agreement between the surrogate model and the target model.

\subsection{Membership Inference}
\label{subsec:method_membership_inference}

\mypara{Threat Model}
The adversary aims to determine whether a given sample is used to train a target model~\cite{SSSS17}.
Following the existing work~\cite{LWHSZBCFZ22,NSH19}, the adversary is assumed to hold a small subset of the training data which is used as member samples\footnote{In many membership inference research, such as~\cite{SSSS17,SZHBFB19,NSH19}, the adversary is assumed to have a shadow dataset to train their shadow model.
To conduct experiments, the researchers split a dataset into four equal parts: two are used as the target model's training and test sets, and the other two are for the shadow model.
We cannot follow the same setting here as the models in \database are all trained.
Thus, we make a stronger assumption for the adversary having access to a partial training set of the target model~\cite{LWHSZBCFZ22,NSH19}.
Such a stronger assumption also allows us to assess the worst-case scenario of membership leakage threat.} and an auxiliary dataset that represents non-member samples.
Note that we use the test set of the corresponding dataset as this auxiliary dataset.
For instance, when experimenting on models trained on ImageNet-1k, the auxiliary dataset is ImageNet-1k's test set.
The adversary also has access to the black-box target model that outputs full posteriors.

\mypara{Methodology}
We use two popular attacks, namely metric-based~\cite{SM21} and MLP-based~\cite{SSSS17,SZHBFB19} attacks.
The former distinguishes member (training) samples from non-member (auxiliary) samples based on behavioral differences in prediction statistics, such as prediction correctness and modified prediction entropy~\cite{SM21}.
For MLP-based attacks, the adversary feeds the member and non-member samples to the target model and gets the output posteriors.
The adversary then trains an attack model (i.e., a binary classifier) based on the posteriors.
In our experiments, the attack model is assembled with one layer of 64 neurons and one layer of 32 neurons using the ReLU activation function.
We train the attack model for 50 epochs using the Adam optimizer with a learning rate of 0.01.

\mypara{Metrics}
We use AUC to evaluate the attack performance.
The higher the AUC, the better the performance is.
Concretely, 0.5 represents random guessing, and 1.0 is a perfect prediction.

\subsection{Backdoor Detection}

In backdoor attacks, the adversary injects the backdoor into the target model without degrading its original performance~\cite{CLLLS17}.
Generally, there are two types of backdoor attacks, i.e., untargeted and targeted.
Untargeted attacks aim to misclassify triggered images, while targeted attacks misclassify triggered images to one specific class.
There exist various types of defenses against backdoors~\cite{WYSLVZZ19,CFZK19,HAS19,GWXDS19,LDG18,LLTMAZ19,CRK19,GXWCRN19,UPWLRC22}.
Since the backdoor injection occurs during training and models in \database are already trained, we choose to evaluate backdoor detection methods on these models.

\mypara{Methodology}
We evaluate three popular backdoor detection methods: Neural Cleanse~\cite{WYSLVZZ19}, Strong Intentional Perturbation (STRIP)~\cite{GXWCRN19}, and NEO~\cite{UPWLRC22}.
The three cover two types of approaches: model inspection and input filtering.

Neural Cleanse is a model inspection approach that aims to detect targeted backdoor attacks.
The key idea of this method is to find the minimal trigger needed to misclassify all samples into each label and leverage an outlier detection method to detect if any trigger candidate is smaller than all the other candidates.
If such an outlier exists, the model is potentially backdoored, and the trigger can be returned for further analysis.
We run Neural Cleanse for 50 epochs and use their default threshold value for detecting outliers.

STRIP is an input-filtering approach.
The key idea is that triggered inputs are less affected by perturbations than normal inputs.
By overlaying various images on the incoming input, the detector examines the randomness (prediction entropy) of the model's prediction on the overlaid input.
The detection considers the model is backdoored if the prediction entropy is low.
We use 2,000 images for testing and the default number of 10 images for overlaying input.

NEO is another input-filtering approach.
The key idea is that triggers within the input contribute the most to the prediction.
The detection first calculates the dominant color of the image, then randomly selects a small region in the input image and replaces it with the dominant color.
If the new prediction differs from the original one, NEO assumes the selected region contains a potential trigger and then superimposes it onto the test set.
If most of the test sets have different predictions after adding the potential trigger, the current input image is labeled as a backdoored image, and the model is also identified as a backdoored model.
We use 200 sampled 4$\times$4 regions (for 32$\times$32 images) and 3 K-means clusters for calculating the dominant color; 0.8 is adopted as the threshold.\footnote{If more than 80\% of the images are misclassified after adding the trigger, the trigger is then confirmed.}
We use 2,000 images from the test set for evaluation.

\mypara{Metrics}
As the models in \database are supposedly free of backdoors, we adopt the false positive rate to evaluate whether these detection methods can falsely recognize clean models as backdoored ones.

\section{Experiment Results}
\label{section:evaluation results}

\begin{figure}[!t]
\centering
\includegraphics[width=0.73\columnwidth]{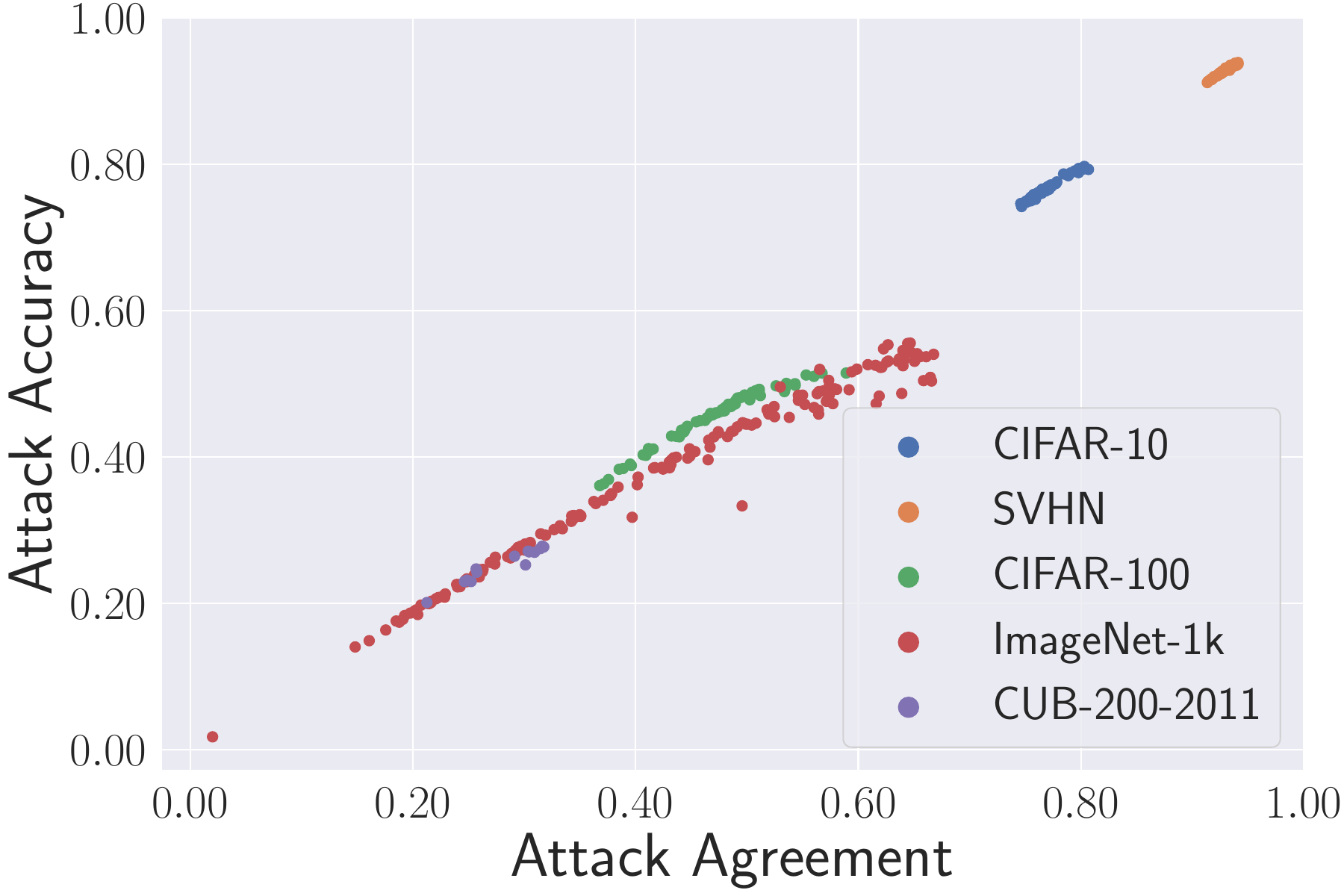}
\caption{The relationship between the attack agreement and the attack accuracy for model stealing on benchmark models across multiple datasets.}
\label{figure:steal_metric}
\end{figure}

\begin{figure*}[!t]
\centering
\begin{subfigure}{0.33\textwidth}
\centering
\includegraphics[width=\textwidth]{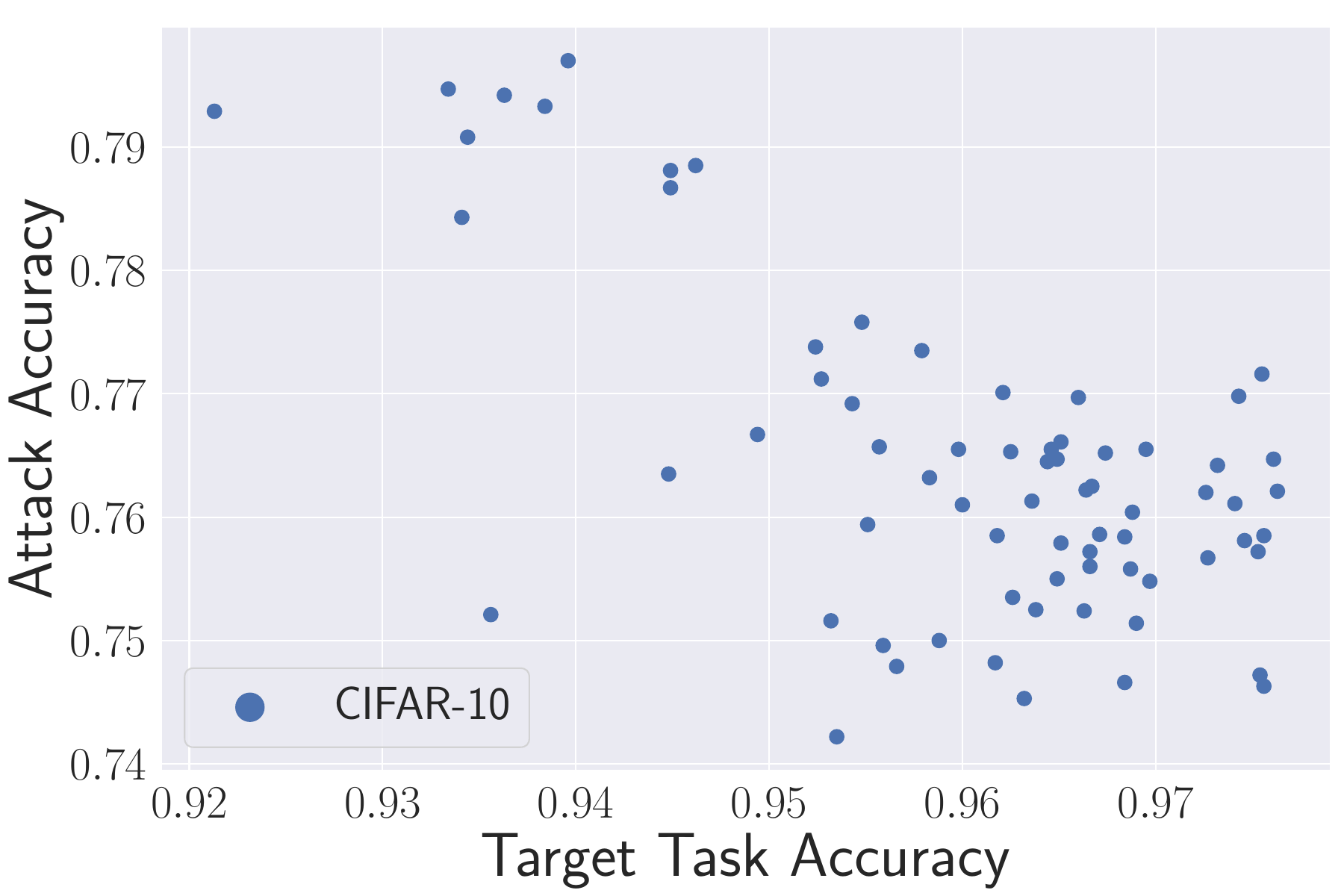}
\caption{CIFAR-10}
\label{figure:steal_cifar10_scatter}
\end{subfigure}%
\hfill
\begin{subfigure}{0.33\textwidth}
\centering
\includegraphics[width=\textwidth]{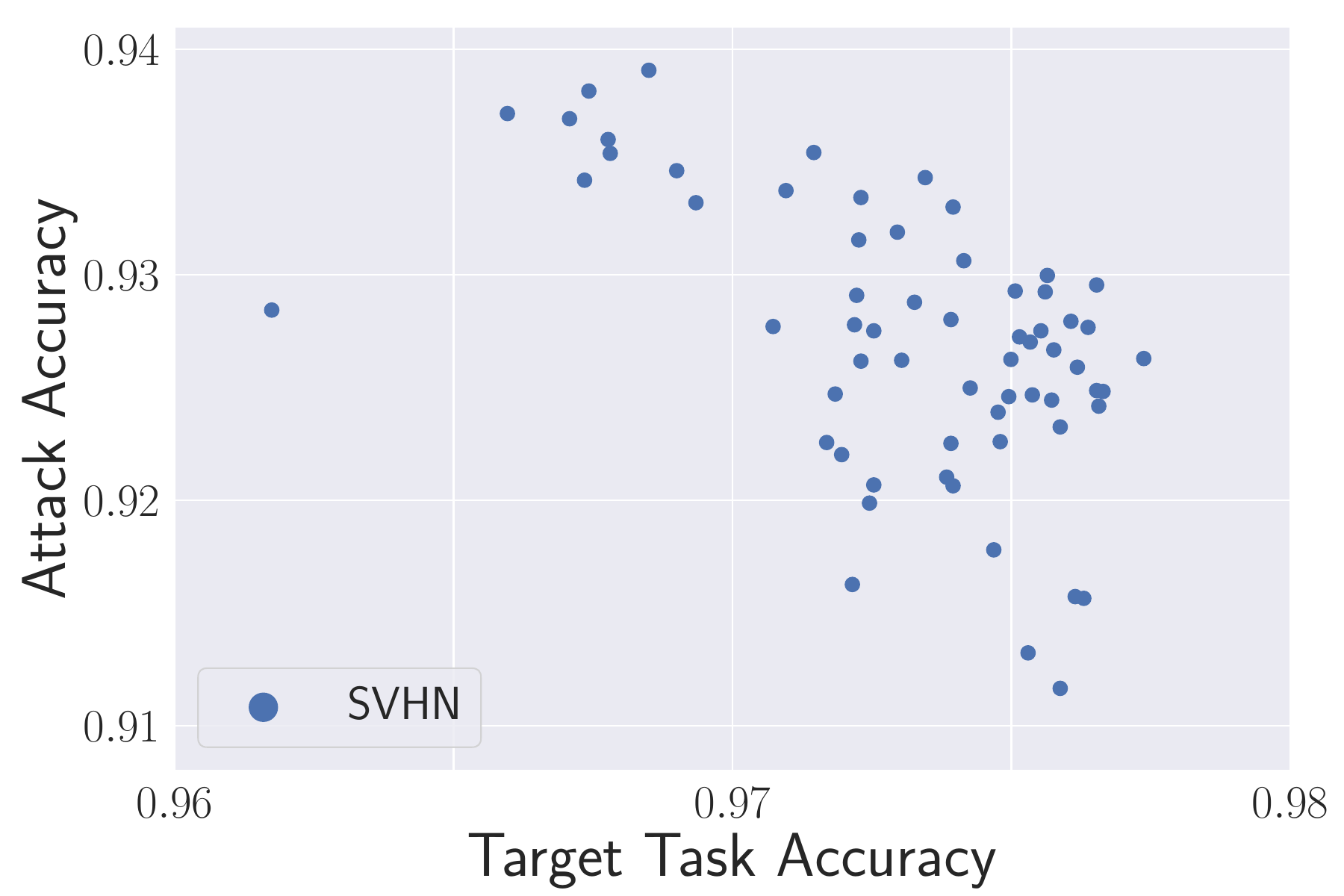}
\caption{SVHN}
\label{figure:steal_svhn_scatter}
\end{subfigure}%
\hfill
\begin{subfigure}{0.33\textwidth}
\centering
\includegraphics[width=\textwidth]{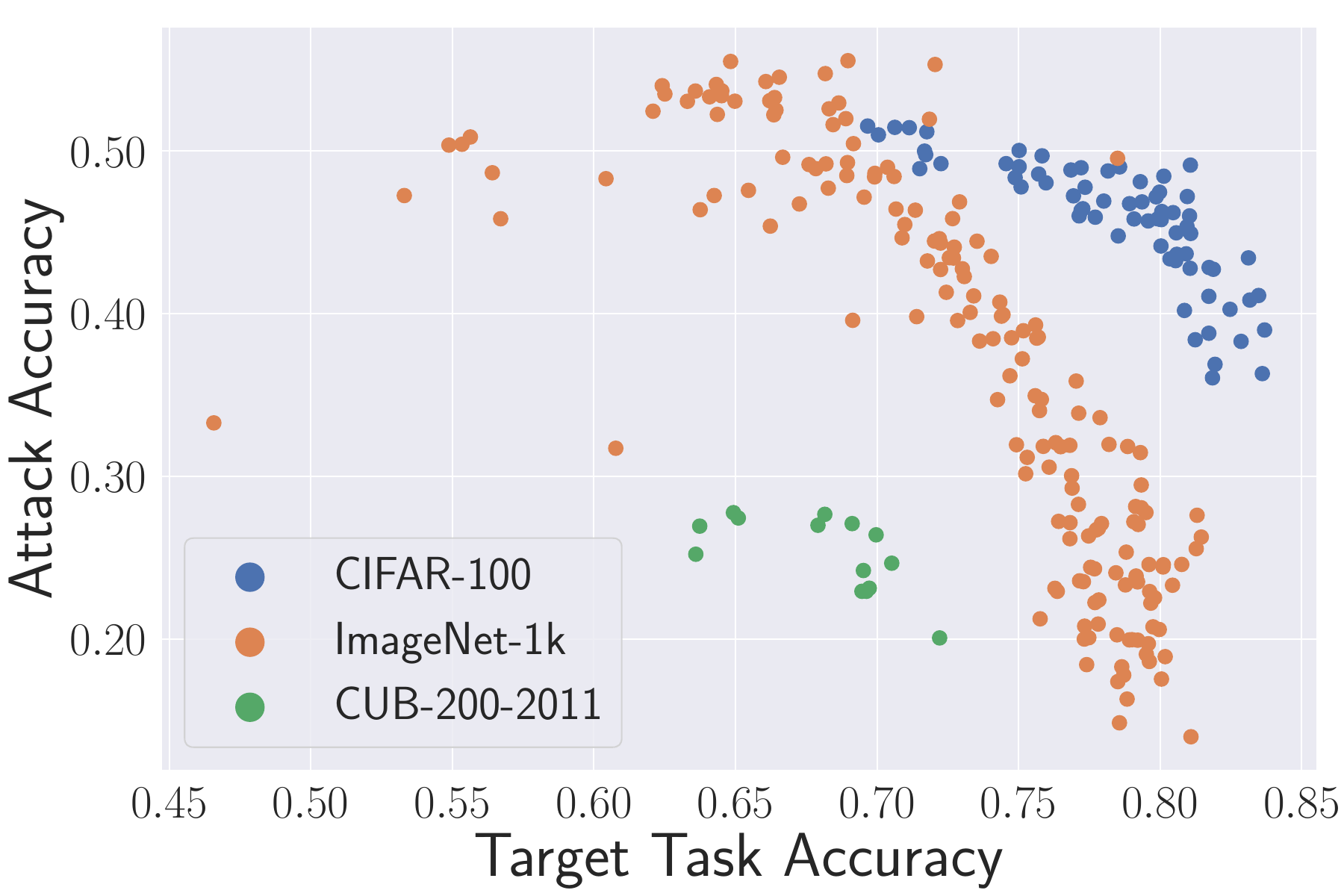}
\caption{Others}
\label{figure:steal_large_scatter}
\end{subfigure}%
\caption{The relationship between the model stealing performance (attack accuracy) and the target model's task accuracy across various benchmark models when using a partial training set as the auxiliary dataset.}
\label{figure:steal_all}
\end{figure*}

\subsection{Model Stealing}
\label{subsec:eval_model_stealing}

We now evaluate the performance of model stealing attacks on public models from \database.

\mypara{The Effect of Target Model's Training Set}
For our evaluation, we primarily use 389 benchmark models trained on CIFAR-10, CIFAR-100, SVHN, ImageNet-1k, and CUB-200-2011 datasets.
These models cover a wide variety of architectures that allow us to make more comprehensive observations on attack behaviors.
First, as seen in \autoref{figure:steal_metric}, we observe a strong positive correlation between the two evaluation metrics (with a 0.991 Pearson correlation coefficient), i.e., the attack agreement (see \autoref{section:methods}) and the attack accuracy on all 389 models.
Due to the page limit, we mainly use attack accuracy as the metric for model stealing in the following analysis.
Secondly, \autoref{figure:steal_all} shows that the attack achieves exceptionally high performance on datasets with a small number of classes and abundant training data, such as CIFAR-10 and SVHN.
For more complex datasets like ImageNet-1k, CIFAR-100, and CUB-200-2011, however, the attack performance significantly deteriorates.
For instance, the average attack accuracy for ImageNet-1k models is 36.3\% while the average target models' accuracy is 73.1\%, i.e., the ratio of the two is 0.496.
Meanwhile, the corresponding ratios for CIFAR-10 and SVHN models are 0.796 and 0.953.
In addition, we find that while the target task performance of CUB-200-2011 models is similar to that of ImageNet-1k models, the attack accuracy on CUB-200-2011 models is significantly lower.\footnote{Note that the model stealing performance on CUB-200-2011 in~\cite{OSF19} is higher than ours, this is because the authors fine-tune their surrogate model and target model on base models pre-trained with ImageNet-1k.}
Note that ImageNet-1k has more classes in total, but CUB-200-2011 has higher class granularity, which means the two datasets have comparable classification complexity.
These results indicate that the model stealing attack is especially ineffective on some outlier datasets, which, to our knowledge, has not been shown previously.

\begin{figure*}[!t]
\centering
\begin{subfigure}{0.33\textwidth}
\centering
\includegraphics[width=\textwidth]{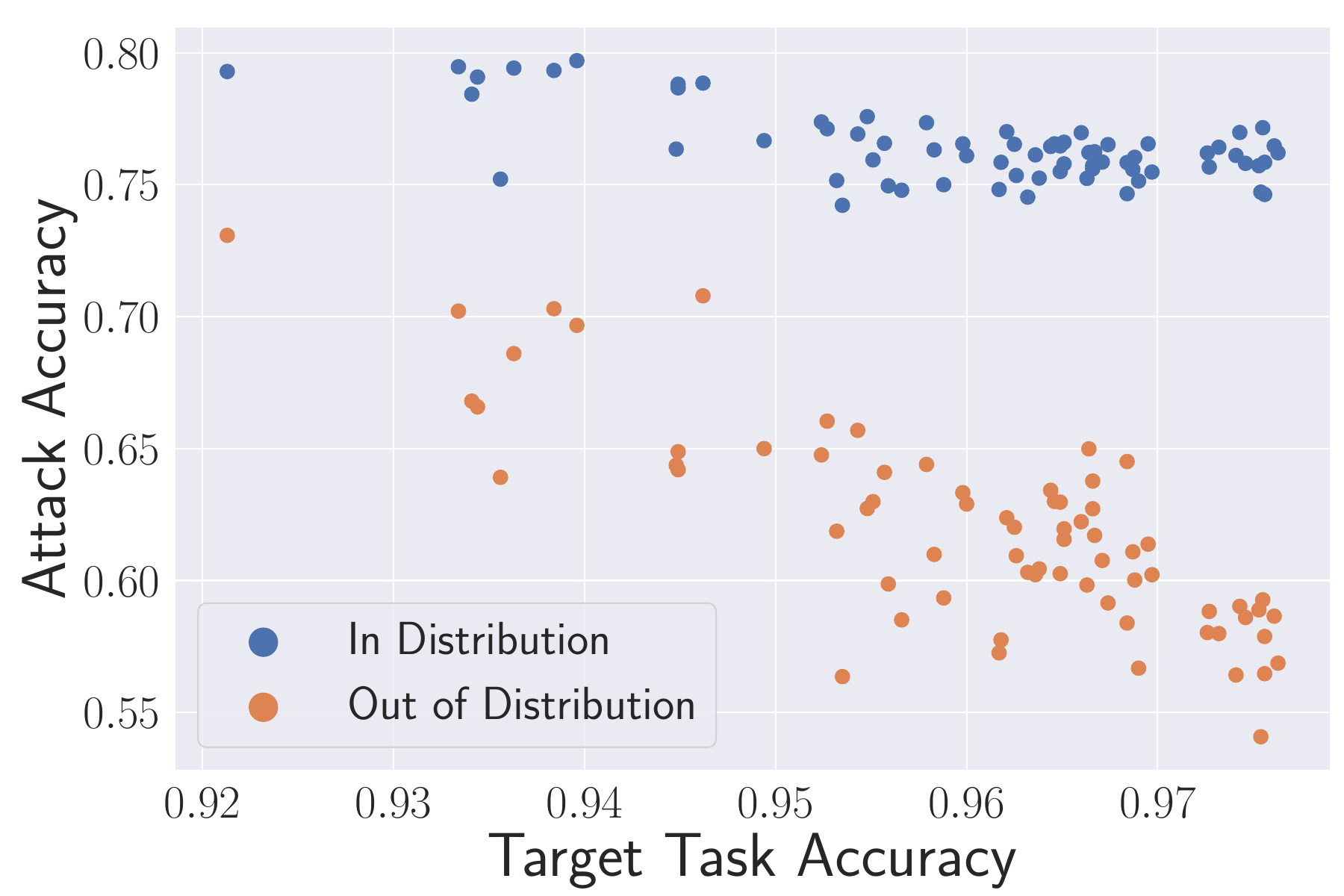}
\caption{CIFAR-10}
\label{figure:cifar10_imgnet_steal}
\end{subfigure}%
\hfill
\begin{subfigure}{0.33\textwidth}
\centering
\includegraphics[width=\textwidth]{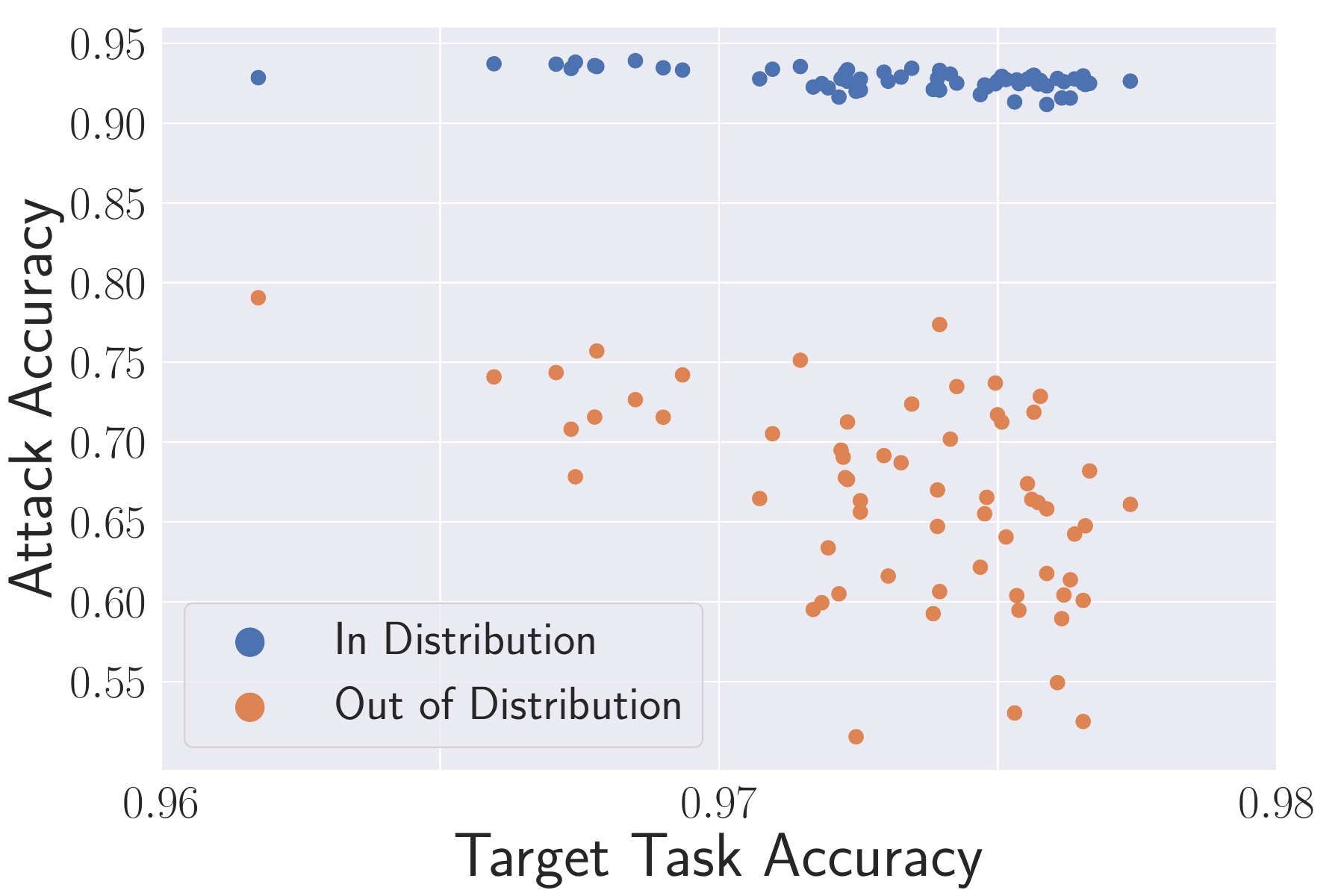}
\caption{SVHN}
\label{figure:svhn_imgnet_steal}
\end{subfigure}%
\hfill
\begin{subfigure}{0.33\textwidth}
\centering
\includegraphics[width=\textwidth]{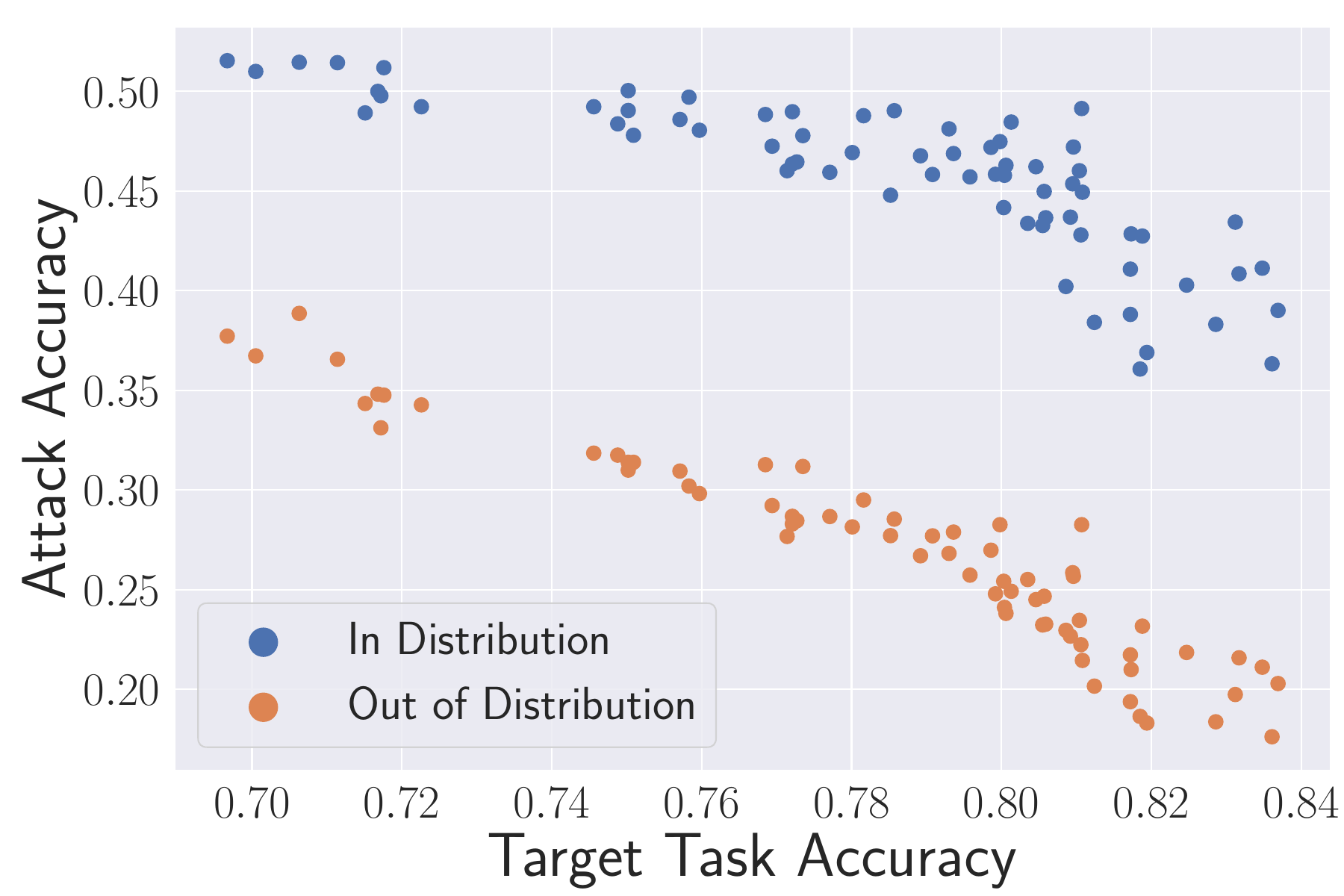}
\caption{CIFAR-100}
\label{figure:cifar100_imgnet_steal}
\end{subfigure}%
\caption{The relationship between the model stealing performance (attack accuracy) and the target model's task accuracy across various benchmark models when using an out-of-distribution auxiliary dataset.}
\label{figure:steal_imgnet_auxiliary}
\end{figure*}

\mypara{Out-Of-Distribution Auxiliary Dataset}
While by default, we assume the adversary has access to a partial training set as the auxiliary dataset, we now consider another scenario where the adversary uses an out-of-distribution dataset to initiate their attack.
More concretely, we leverage a subset of the large and diverse ImageNet-1k dataset as the auxiliary dataset to steal benchmark models trained on CIFAR-10, SVHN, CIFAR-100, and CUB-200-2011.
\autoref{figure:steal_imgnet_auxiliary} shows the attack performance deteriorates across a large number of benchmark models when using the out-of-distribution auxiliary dataset.
In addition, we find that the attack performance on SVHN models decreases more significantly than on CIFAR-10 models.
For instance, the average attack accuracy decreases by 27.8\% on SVHN and 19.0\% on CIFAR-10, respectively.
We suspect this is due to the fact that the images in ImageNet-1k are more similar to the images in CIFAR-10 than to the images in SVHN.\footnote{Previous work~\cite{OSF19} makes similar observations.}
Further, on a more complex dataset, the attack performance can suffer even greater, e.g., with an average degradation of 41.7\% on CIFAR-100.
For the previous poor-performing CUB-200-2011 models, the attack accuracy also decreases by 21.4\%, even though the auxiliary dataset contains many classes similar to the original dataset (e.g., bird species), and overall has more samples than the original partial training set (100k vs.\ 5k).
In summary, we find that using out-of-distribution data as the auxiliary dataset does not benefit model stealing.

\begin{figure*}[!t]
\centering
\begin{subfigure}{0.33\textwidth}
\centering
\includegraphics[width=\textwidth]{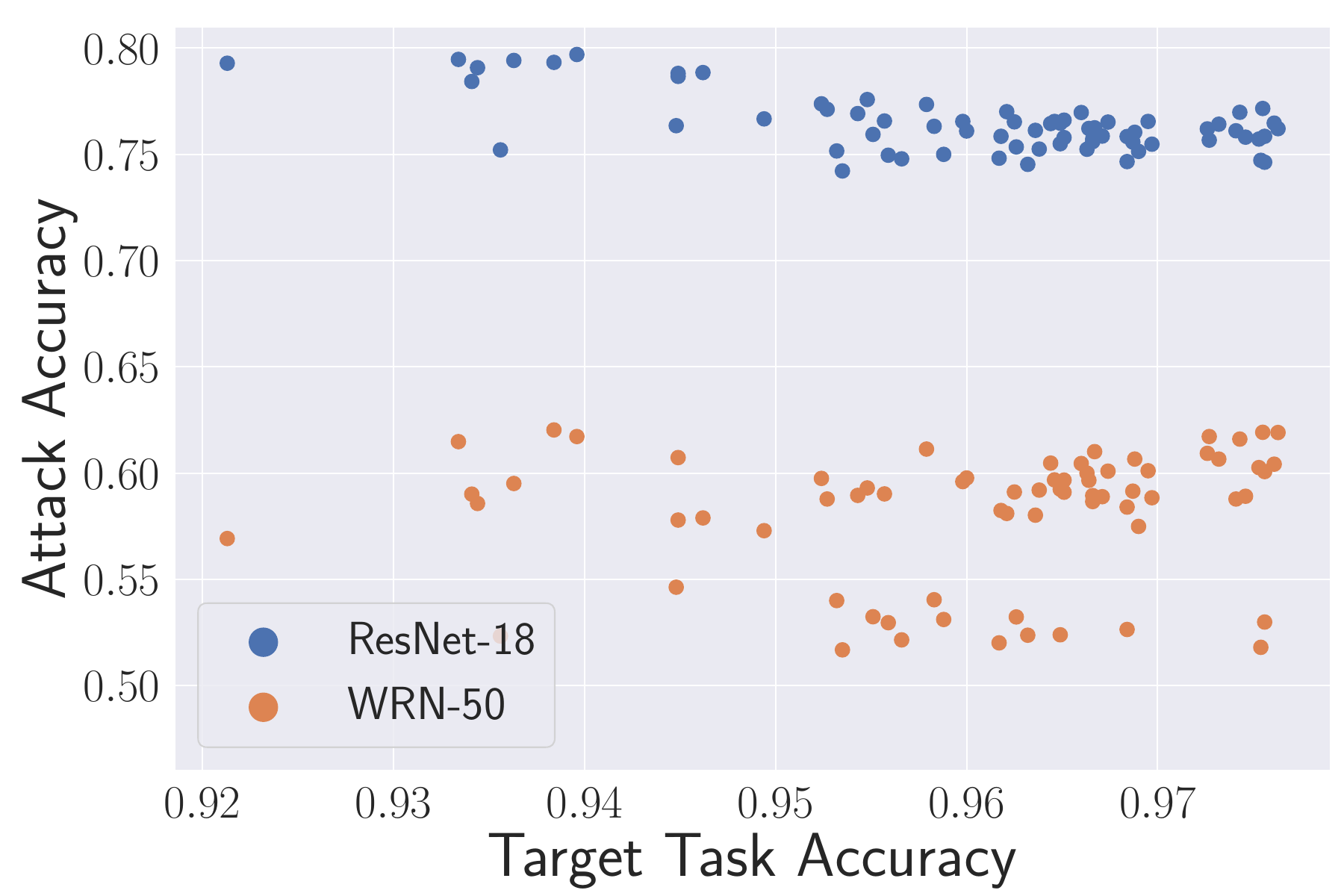}
\caption{CIFAR-10}
\label{figure:cifar10_wrn_steal}
\end{subfigure}%
\hfill
\begin{subfigure}{0.33\textwidth}
\centering
\includegraphics[width=\textwidth]{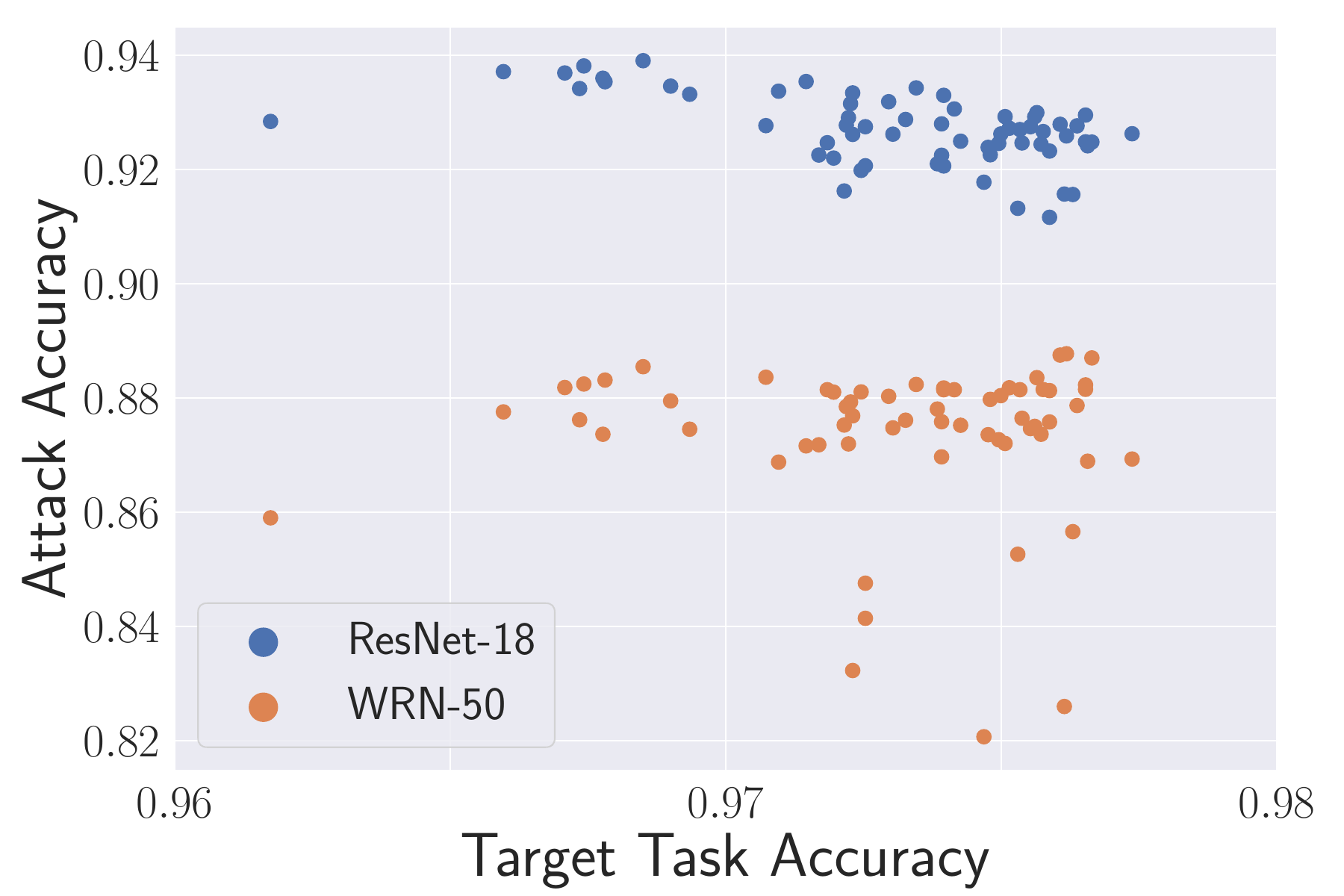}
\caption{SVHN}
\label{figure:svhn_wrn_steal}
\end{subfigure}%
\hfill
\begin{subfigure}{0.33\textwidth}
\centering
\includegraphics[width=\textwidth]{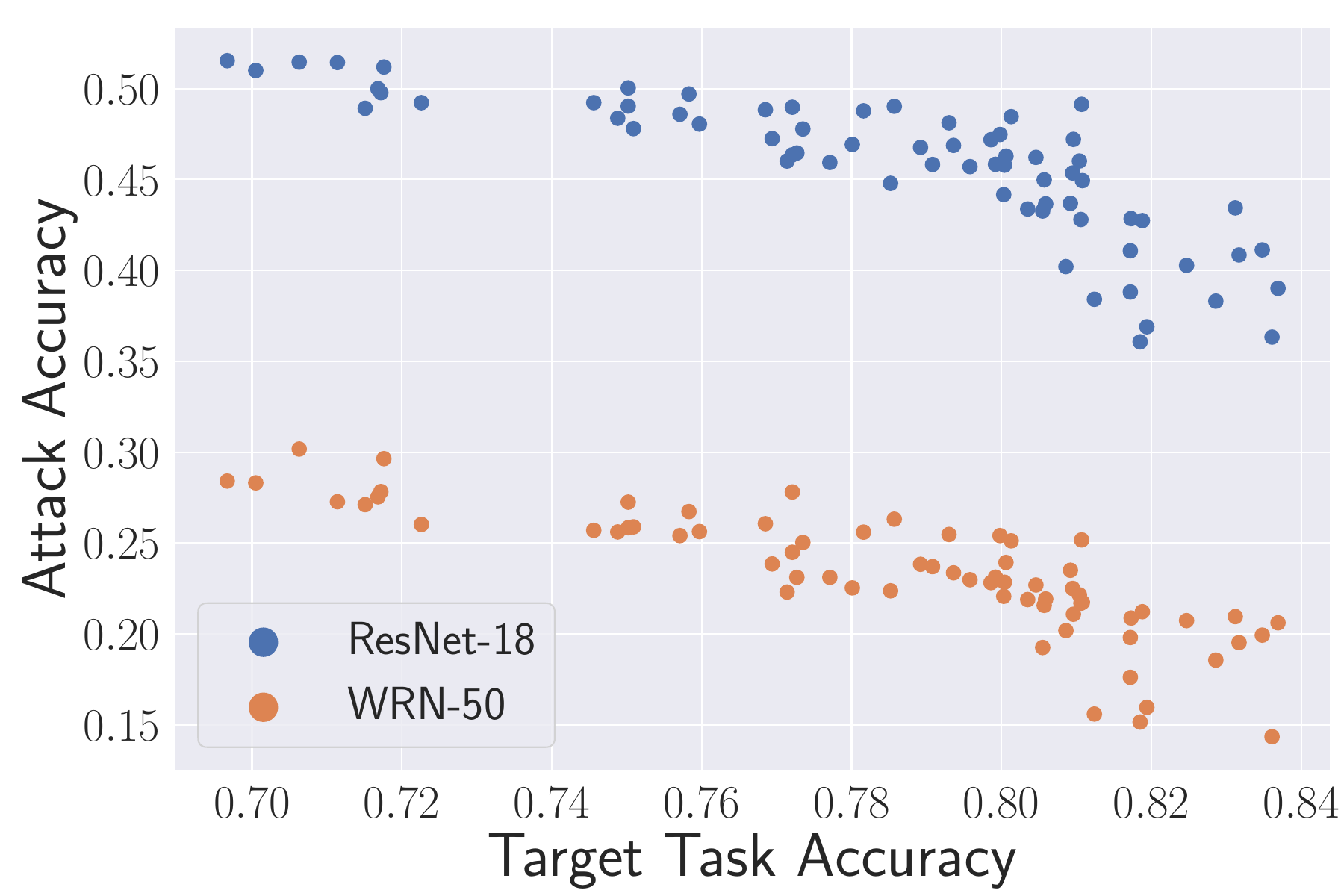}
\caption{CIFAR-100}
\label{figure:cifar100_wrn_steal}
\end{subfigure}%
\caption{The relationship between the model stealing performance (attack accuracy) and the target model's task accuracy across various benchmark models when using a more complex surrogate model, i.e., WRN-50.}
\label{figure:steal_wrn}
\end{figure*}

\mypara{The Effect of Target Model's Performance}
From the perspective of target models' inherent properties, we mainly study the relation between their target task performance and the corresponding model stealing attacks' performance.
\autoref{figure:steal_all} shows a distinct negative correlation between the two.
While the correlation is clear on both smaller datasets, such as CIFAR-10 (-0.693) and SVHN (-0.603), it is more evident on larger and more complex datasets, like ImageNet-1k (-0.844) and CIFAR-100 (-0.873).
We present the Pearson correlation coefficients in the parenthesis.
This negative correlation has been observed by Jagielsk et al.~\cite{JCBKP20} previously.
Our finding, however, differs in magnitude.
Due to the wide range and variety of benchmark models, we find that the attack is largely likely to fail on high-performing models.
For example, on most ImageNet-1k models with target task accuracy above 75\%, the attack accuracy does not surpass 40\%.
In a more concrete example, our model stealing attack on RegNetY-320~\cite{RKGHD20} with 146 million parameters only achieves 25.6\% attack accuracy.
This implies that the model stealing performing well on simpler architectures does not guarantee success on complex and high-performing models.

\mypara{More Complex Surrogate Model}
All results above are done using the ResNet-18 architecture as the surrogate model.
Previous work~\cite{OSF19} has shown that using a more complex surrogate model can improve model stealing performance.
We now conduct the attack with a larger surrogate model, i.e., WRN-50~\cite{ZK16}, to evaluate whether similar observations can be made on benchmark models as well.
\autoref{figure:steal_wrn} shows the attack performance actually deteriorates on all three datasets.
Specifically, for the SVHN, CIFAR-10, and CIFAR-100 models, the attack accuracy degrades by an average of 5.8\%, 29.0\%, and 47.3\%, respectively.
In contrast to previous work, our experiments show that larger and more complex surrogate models do not improve model stealing performance on \database.

\mypara{Benchmark vs.\ Security Models}
As mentioned in \autoref{section:database}, we also extensively search for public models used in security/privacy research (named security models).
Here, we examine whether the attack behaves differently on security models compared to the benchmark models above.

\begin{figure}[!t]
\centering
\includegraphics[width=0.73\columnwidth]{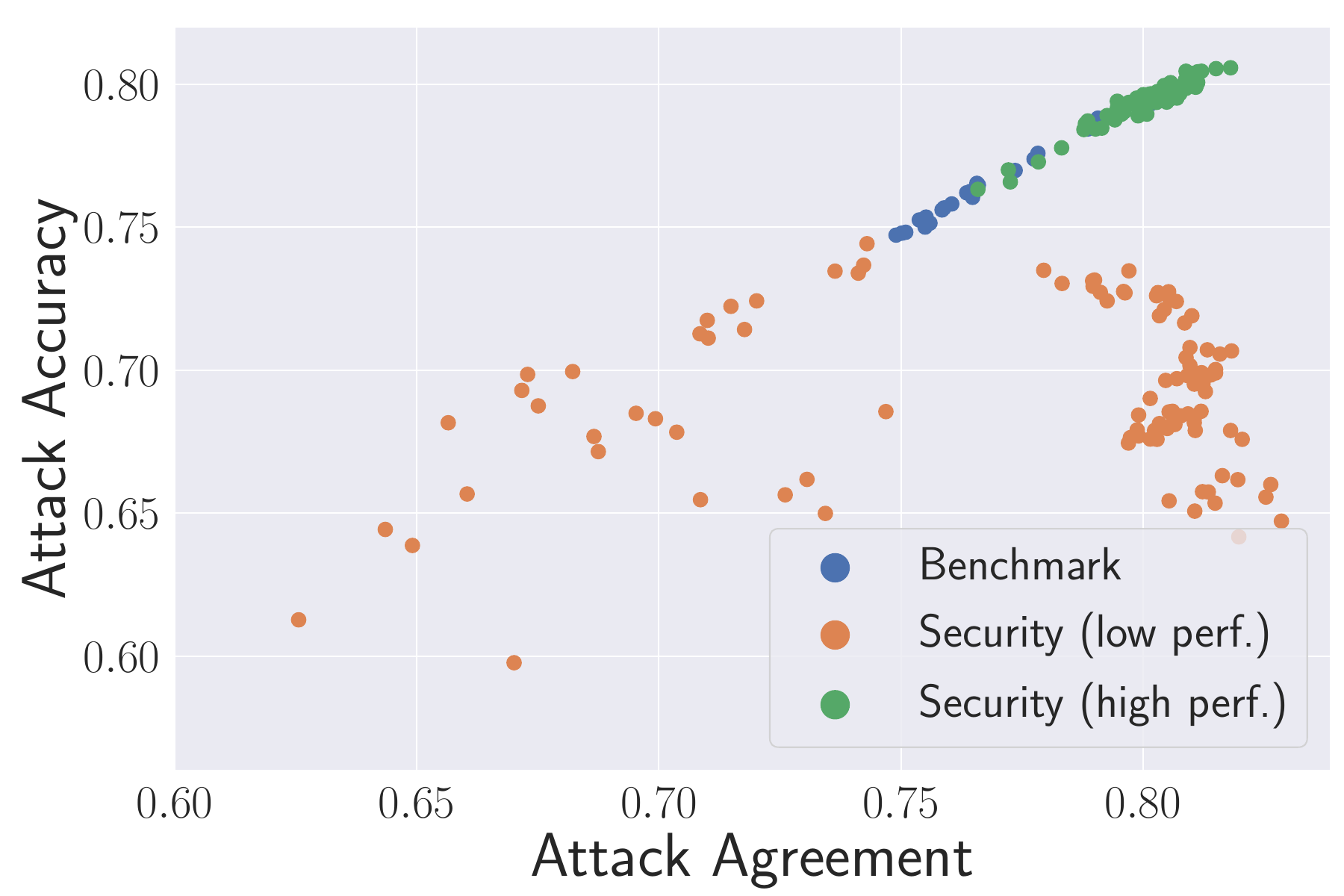}
\caption{The relationship between the attack agreement and the attack accuracy for model stealing on CIFAR-10 benchmark and security models.}
\label{figure:steal_metric_adv}
\end{figure}

\begin{figure}[!t]
\centering
\includegraphics[width=0.73\columnwidth]{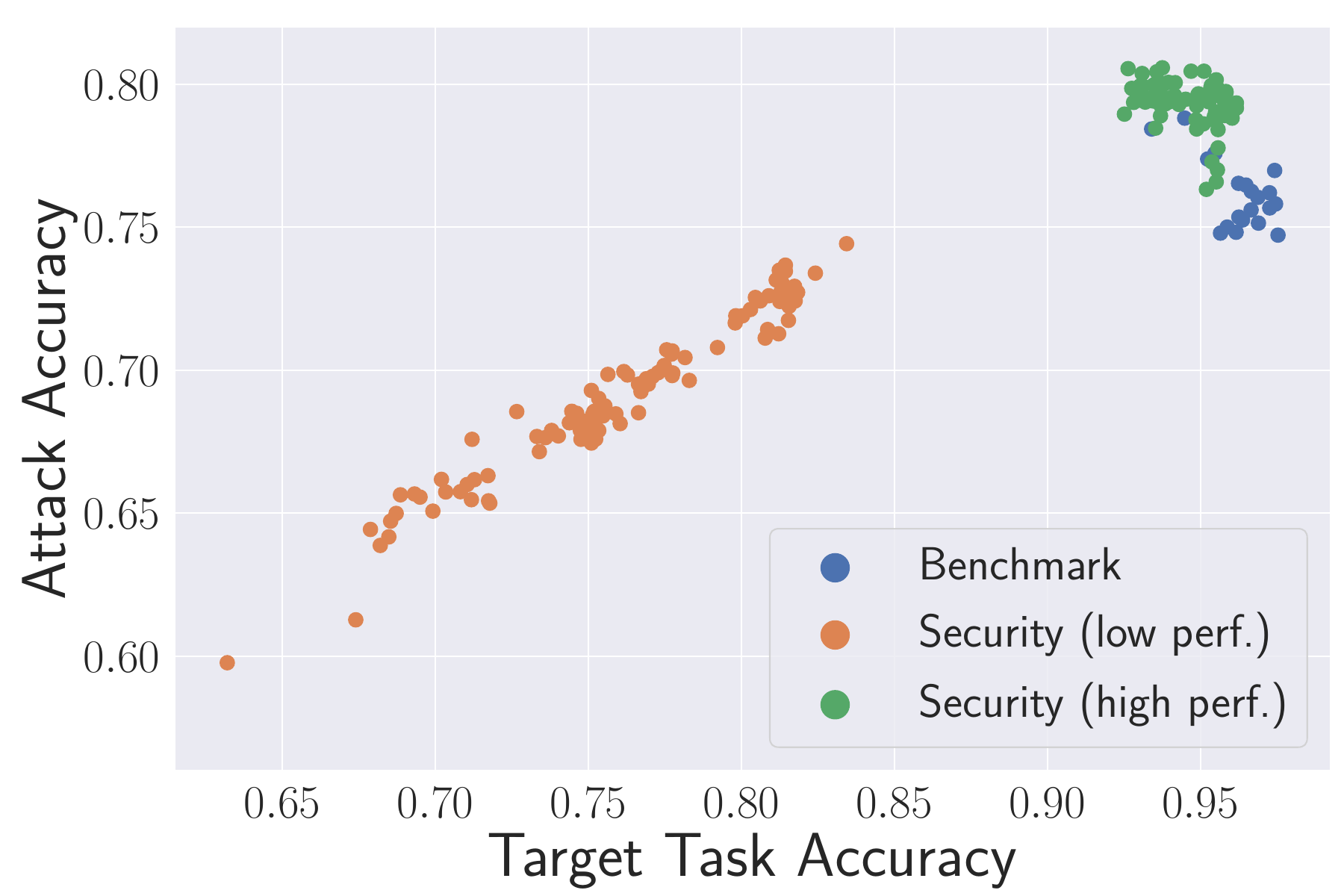}
\caption{The relationship between the model stealing performance (attack accuracy) and the target model's task accuracy on CIFAR-10 benchmark and security models.}
\label{figure:steal_acc_adv}
\end{figure}

Recall that the security models trained on CIFAR-10 can be divided into two clusters in terms of target task performance (see \autoref{figure:violin_target_accuracy}).
Both \autoref{figure:steal_metric_adv} and \autoref{figure:steal_acc_adv} show the cluster of high-performing security models behave very similarly to the benchmark models, where the two evaluation metrics have a high agreement, and the attack performance is negatively correlated (-0.362) with model's target task performance.
Interestingly, the low-performing cluster shows drastically different behavior.
First, many of these models have high attack agreement while the attack accuracy varies.
Secondly, the correlation between the attack accuracy and target task accuracy is distinctively positive (0.998).
We find models used for studying model stealing attacks in previous works~\cite{LWHSZBCFZ22} exhibit similar behavior.\footnote{We can infer the same positive correlation from the negative correlation between their models' overfitting level (the difference between training and test accuracy) and the model stealing performance since all of their models have 100\% training accuracy.}
This drastic change in correlation indicates that when the target model is not trained to its maximum ``potential,'' the attack can behave differently.
For future research on the security and privacy of machine learning models, we advise the researchers to use benchmark models when possible or train the target models to high performance.

\begin{figure}[!t]
\centering
\includegraphics[width=0.73\columnwidth]{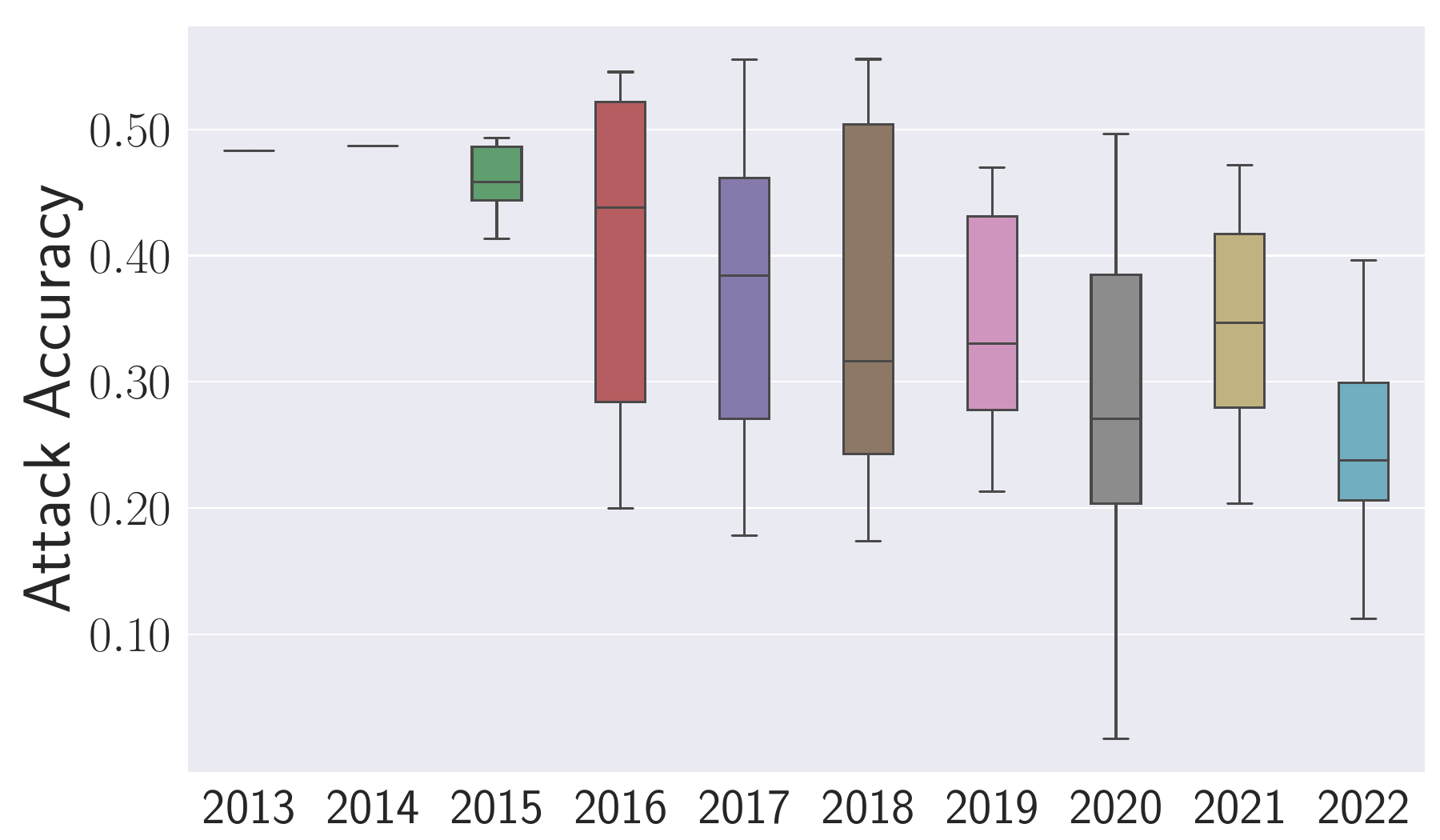}
\caption{The model stealing performance (attack accuracy) with respect to the publication year.}
\label{figure:modsteal_year}
\end{figure}

\begin{figure}[!t]
\centering
\includegraphics[width=0.73\columnwidth]{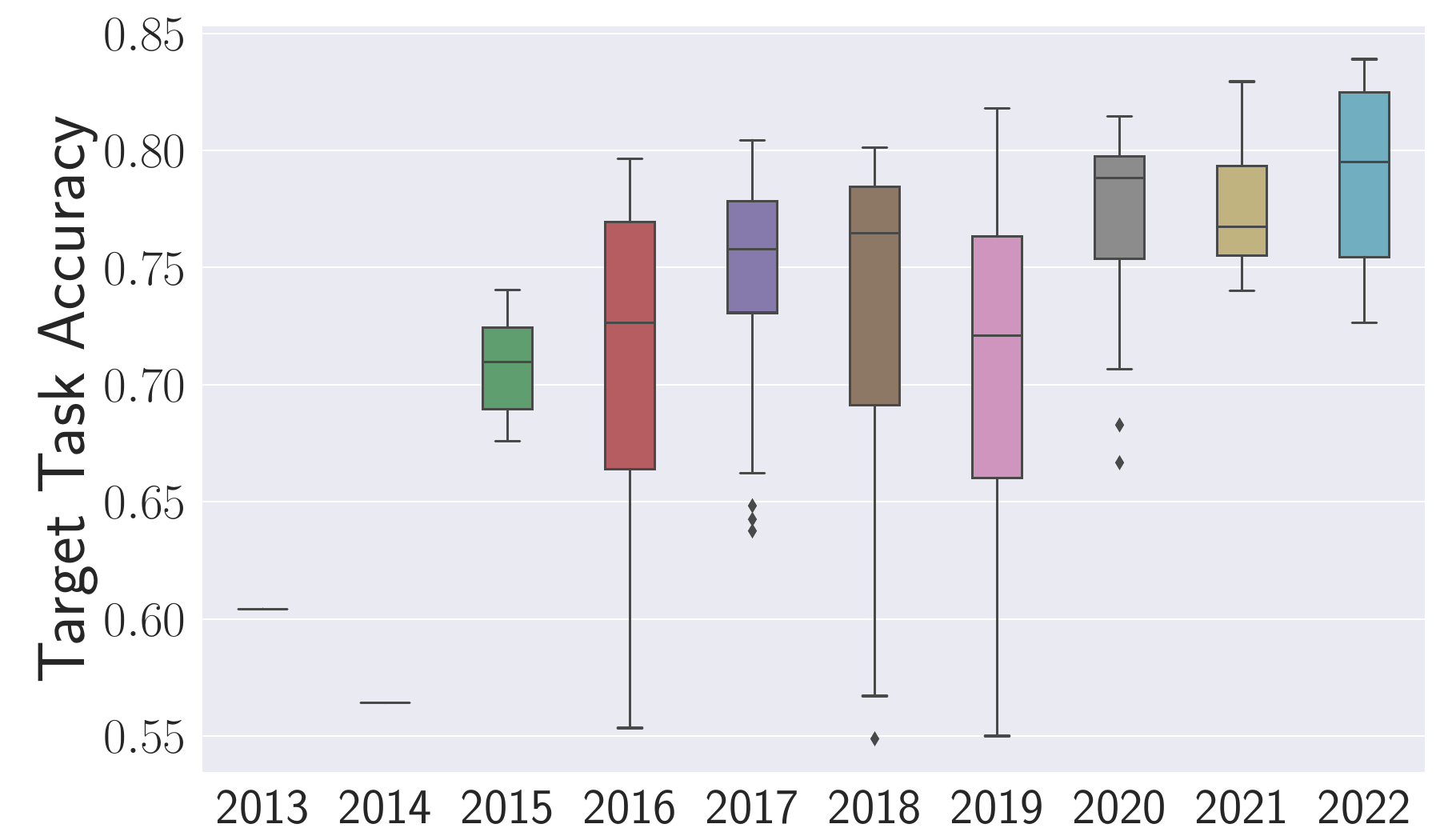}
\caption{The model's target task performance with respect to the publication year.}
\label{figure:acc_year}
\end{figure}

\begin{figure}[!t]
\centering
\includegraphics[width=0.73\columnwidth]{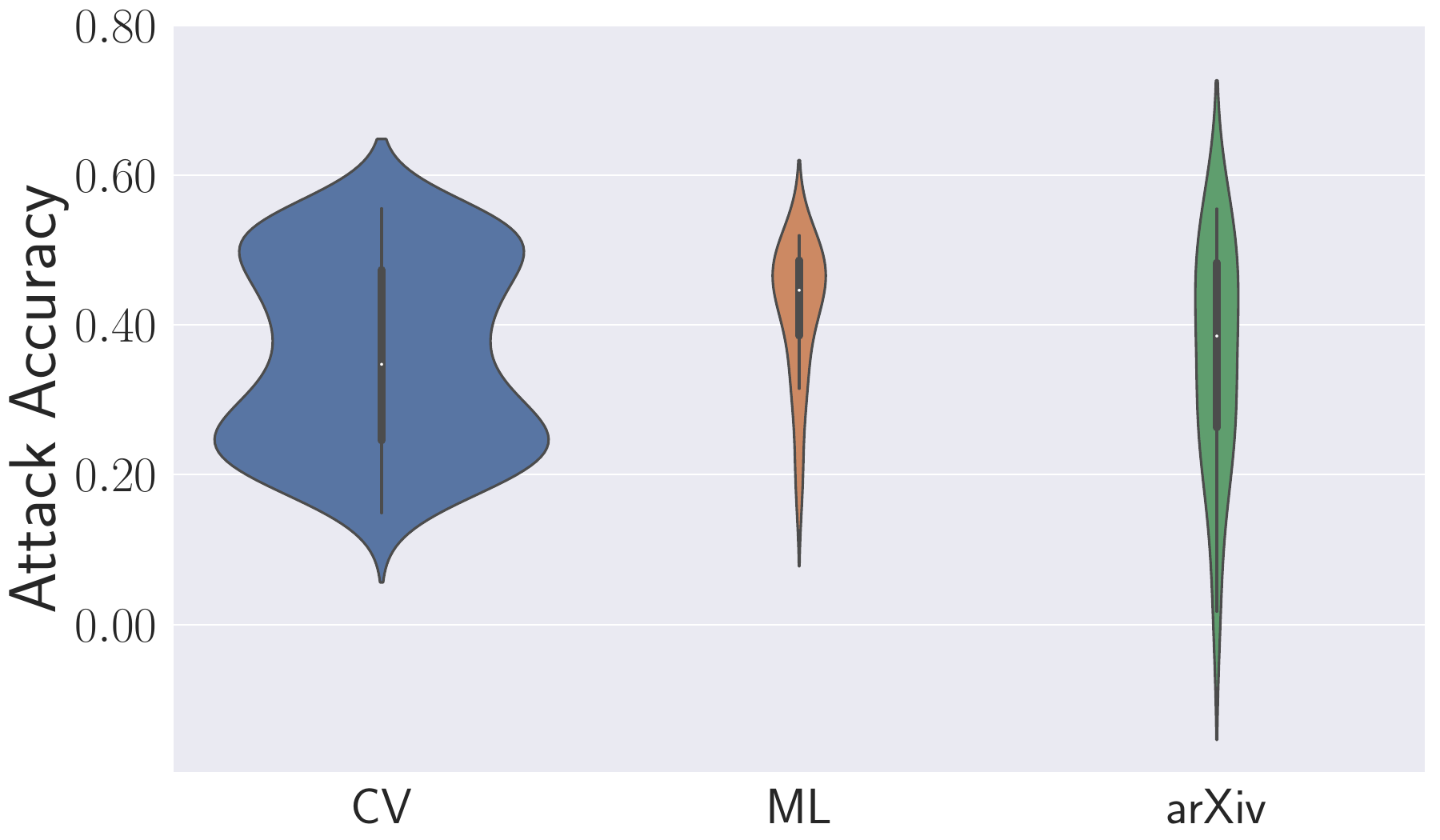}
\caption{The model stealing performance (attack accuracy) with respect to the conference type.}
\label{figure:modsteal_conftyp}
\end{figure}

\mypara{Metadata}
Next, we examine how some metadata relates to the attack performance using public models trained on ImageNet-1k.
On the time dimension, \autoref{figure:modsteal_year} shows the model stealing attack is generally less effective on newer models, which may be due to the higher performance of the newer models, see \autoref{figure:acc_year}.
Besides, as the violin plot shows in \autoref{figure:modsteal_conftyp}, the distribution of the CV domain is much wider than that of other domains, which means that computer vision conferences are the more popular venue for publishing new model architectures.
However, the attack shows no significant difference between different types of publishing venues.

\subsection{Membership Inference}

We next evaluate the performance of membership inference attacks on public models from \database.

\begin{figure*}[!t]
\centering
\begin{subfigure}{0.33\textwidth}
\centering
\includegraphics[width=\textwidth]{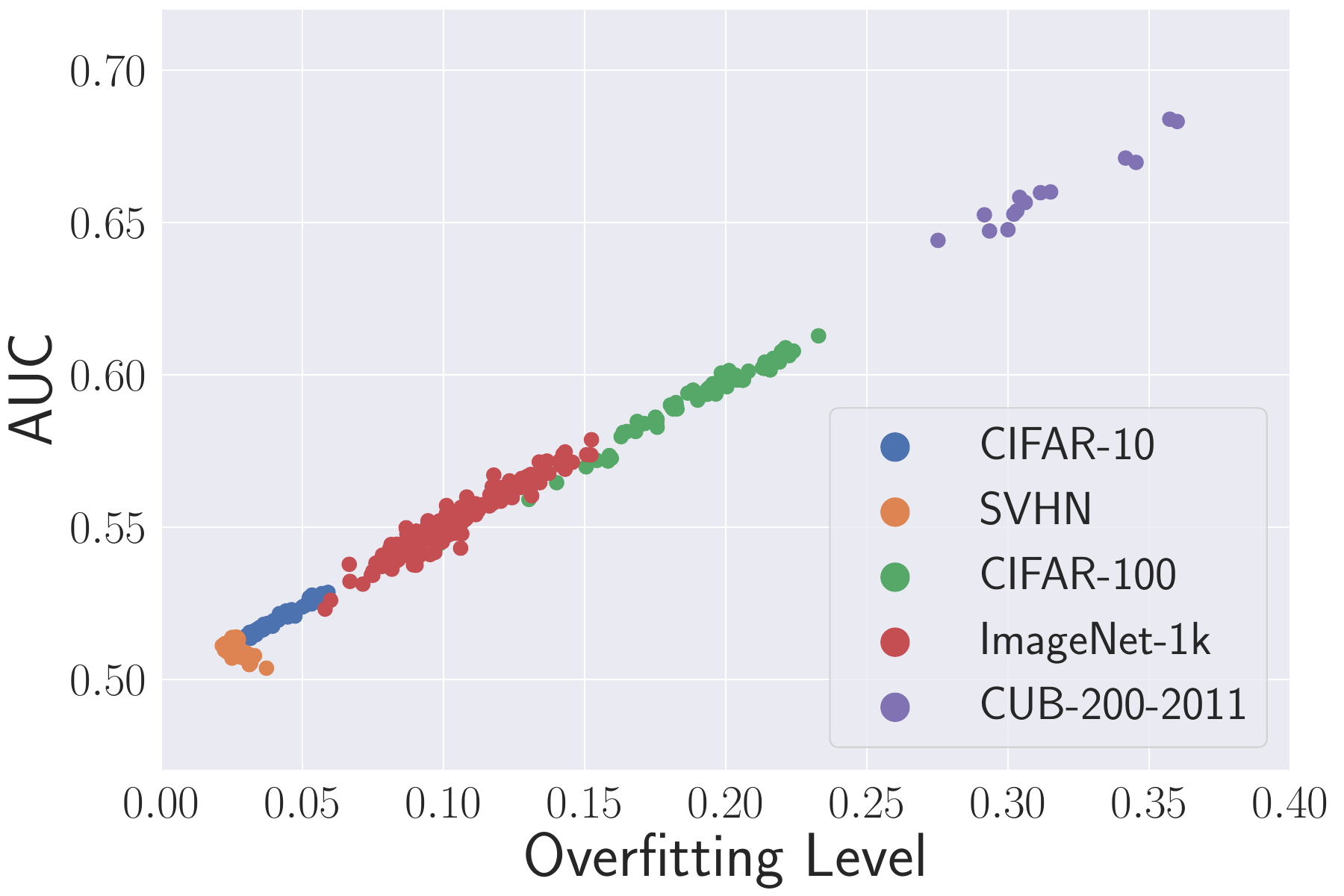}
\caption{Prediction Correctness}
\label{figure:correct_overfit_all}
\end{subfigure}%
\hfill
\begin{subfigure}{0.33\textwidth}
\centering
\includegraphics[width=\textwidth]{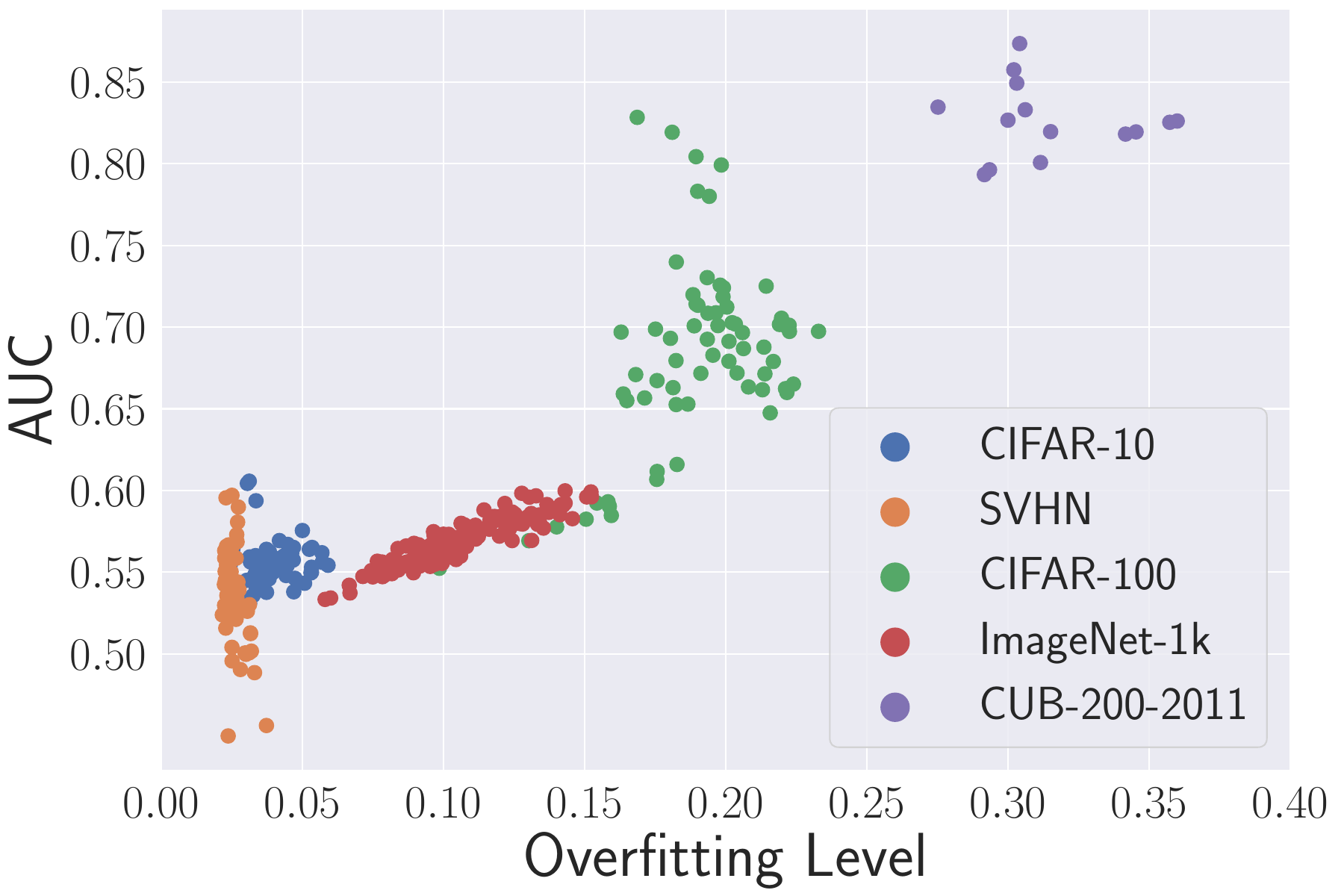}
\caption{Modified Entropy}
\label{figure:mentropy_overfit_all}
\end{subfigure}%
\hfill
\begin{subfigure}{0.33\textwidth}
\centering
\includegraphics[width=\textwidth]{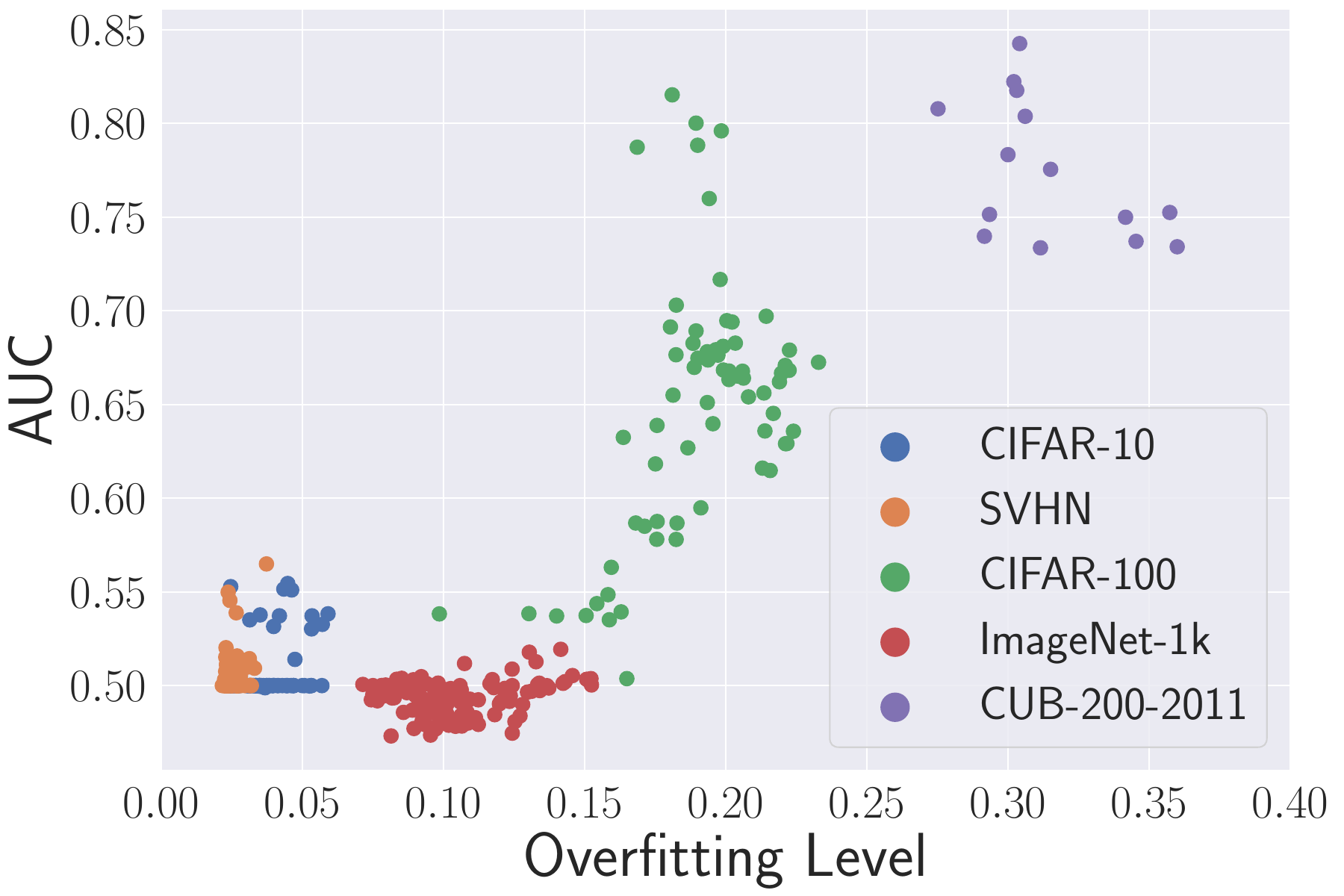}
\caption{MLP Full Posterior}
\label{figure:mlp_overfit_all}
\end{subfigure}%
\caption{The performance (AUC) of different membership inference attacks with respect to the target model's overfitting level.}
\label{figure:membership_all}
\end{figure*}

\mypara{The Effect of Target Model's Overfitting Level}
Similar to the previous section, we compare the attack performance on benchmark models trained on several different datasets.
First of all, we evaluate the membership inference attack performance on benchmark models with respect to their overfitting level.
Here, the overfitting level means the difference between training and test accuracy on models' original datasets~\cite{LWHSZBCFZ22}.
As shown in \autoref{figure:membership_all}, we make similar observations as in many previous works~\cite{SSSS17,LWHSZBCFZ22}, where membership inference achieves better performance on victim models with a higher overfitting level.
We can still find such an association even when the overall attack is not very effective.
For instance, the correlation is still present on CIFAR-10 (0.204) and ImageNet-1k (0.247) models, even though the AUC is generally lower than 0.6.
This correlation is expected since the attack relies heavily on the different distributions of posterior probabilities between member and non-member samples.

\begin{figure*}[!t]
\centering
\begin{subfigure}{0.33\textwidth}
\centering
\includegraphics[width=\textwidth]{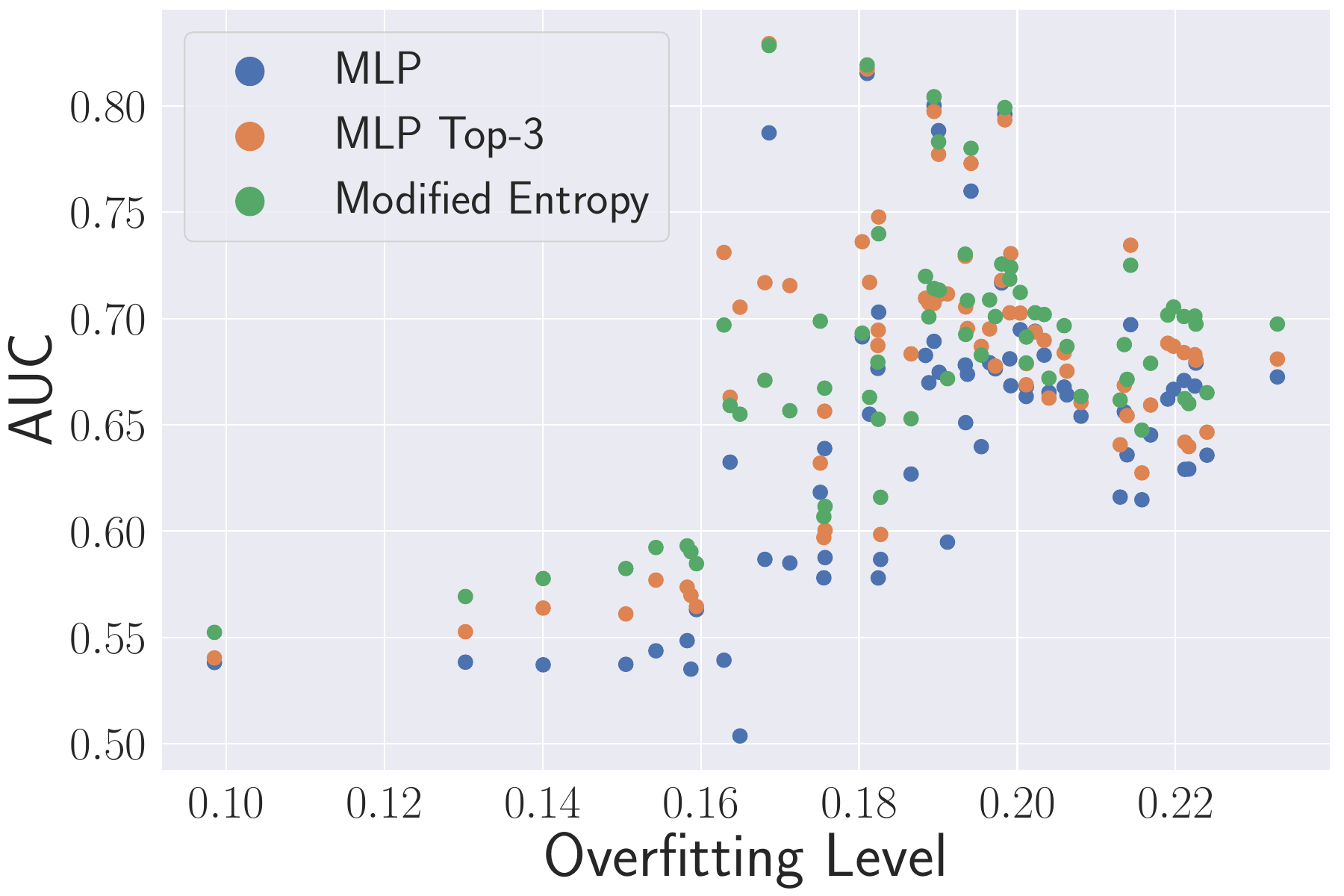}
\caption{CIFAR-100 Models}
\label{figure:cifar100_top3}
\end{subfigure}%
\hfill
\begin{subfigure}{0.33\textwidth}
\centering
\includegraphics[width=\textwidth]{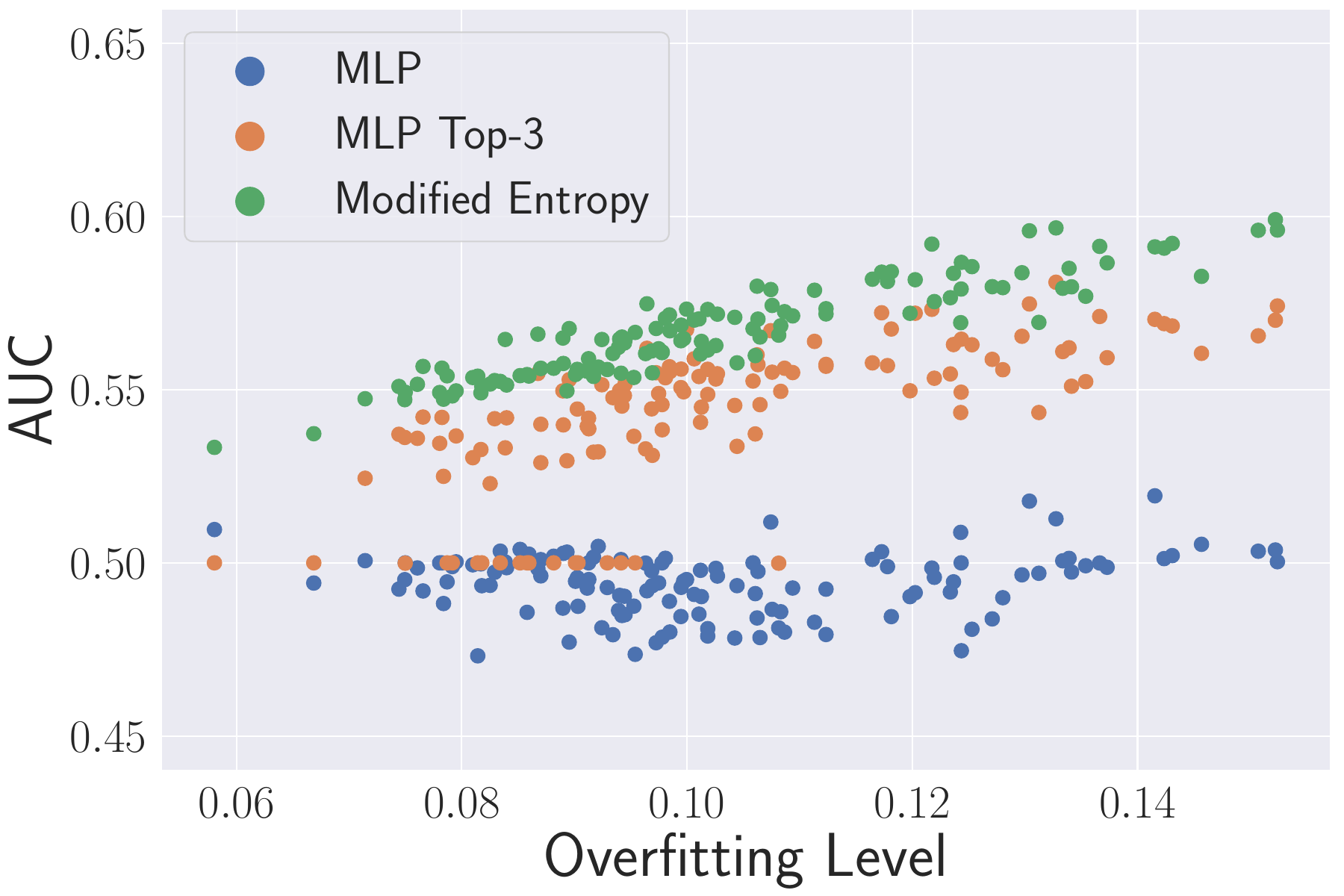}
\caption{ImageNet-1k Models}
\label{figure:imagenet_top3}
\end{subfigure}%
\hfill
\begin{subfigure}{0.33\textwidth}
\centering
\includegraphics[width=\textwidth]{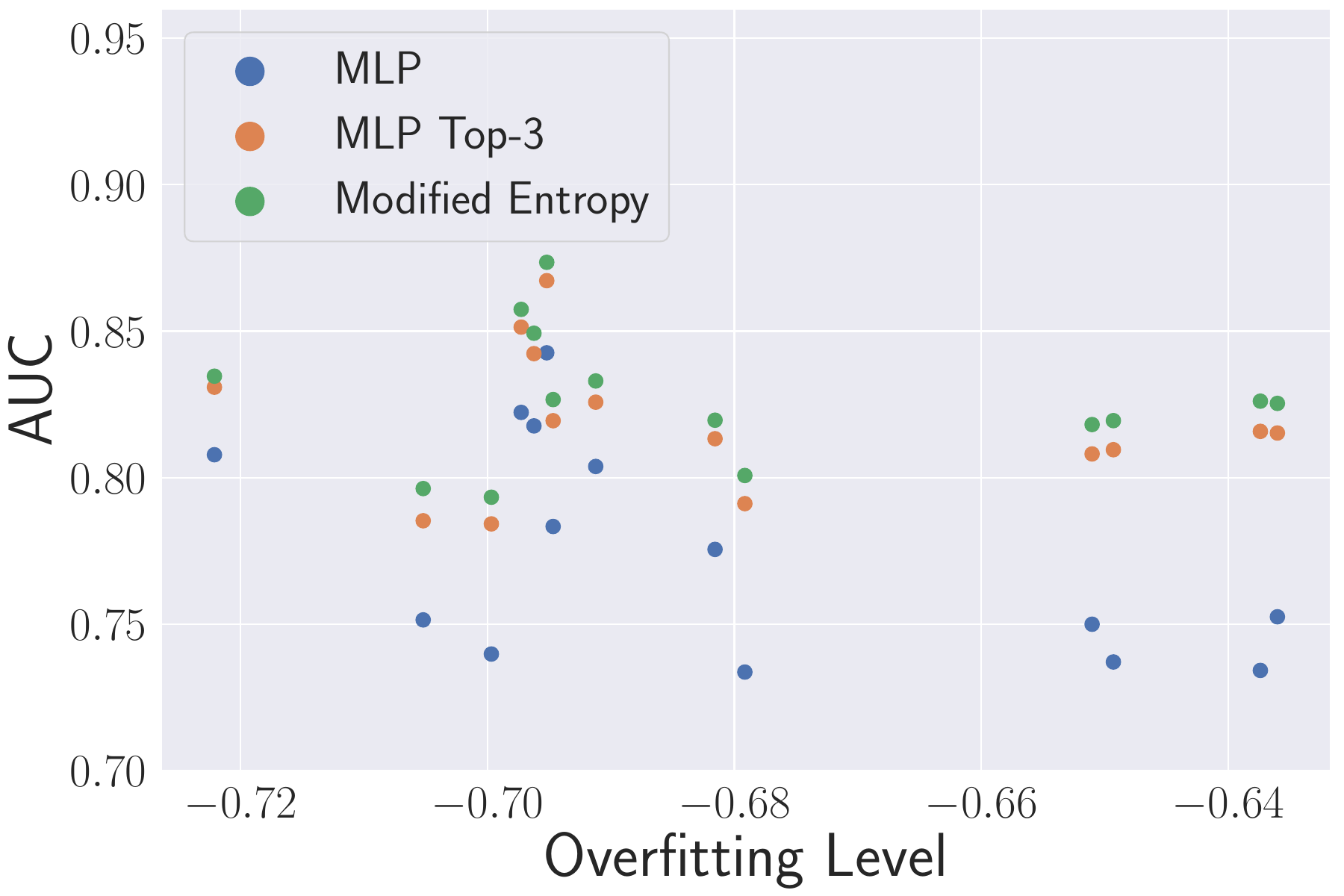}
\caption{CUB-200-2011 Models}
\label{figure:cub_top3}
\end{subfigure}%
\caption{The performance (AUC) of MLP-based and metric-based membership inference attacks with respect to the target model's overfitting level.}
\label{figure:membership_mlp_methods}
\end{figure*}

\mypara{The Effect of Target Model's Training Set}
We also compare the membership inference attack performance from the perspective of the target model's training set.
For models trained on simpler datasets with a small number of classes and sufficient training data, the attack performs poorly, as seen in \autoref{figure:membership_all}.
More concretely, for the SVHN and CIFAR-10 models, the average attack performance is only slightly better than random guessing, with an AUC around 0.565 for the modified entropy attack, and not better than random guessing in most cases for the MLP-based attack.
However, many previous works~\cite{SSSS17,SZHBFB19} evaluate their attack performance using these two datasets and achieve much higher performance than random guessing, which is quite inconsistent with our observations on benchmark models.
This is likely due to the higher prediction accuracy and the lower overfitting level of the benchmark models in \database compared to the self-trained ones in previous works.

For more complex datasets, e.g., CIFAR-100 and CUB-200-2011, the attack performance is significantly better than the previous two simple datasets.
For instance, the modified entropy attack can achieve higher than 0.8 AUC on almost all CUB-200-2011 models.
These results indicate that even these heavily fine-tuned benchmark models cannot achieve the low overfitting level as the ones on simpler datasets.

Interestingly, the membership inference attack does not achieve high performance on the benchmark ImageNet-1k models, even though the overfitting level is not as low as that of the ones trained on simpler datasets.
We suspect the large and diverse training set for these benchmark models makes the task of membership inference more difficult.
To our knowledge, this observation has not been made previously.

\mypara{Different Attack Methods' Effectiveness}
As mentioned in \autoref{subsec:method_membership_inference}, we evaluate three attack methods, namely prediction correctness (metric-based) attack, modified entropy (metric-based) attack, and MLP-based attack.
Our experiments demonstrate that the modified entropy attack and MLP-based attack show varying degrees of success and correlation with overfitting level, depending on the training set of the victim model, as previously seen in \autoref{figure:membership_all}.
The prediction correctness attack, unlike the other two, strictly follows the model's overfitting level across all datasets evaluated,\footnote{SVHN models' overfitting levels are too low for the attack to be stable.} which implies its better transferability to unknown models and datasets.
This method, however, achieves only limited success even on models with high overfitting levels and generally performs worse than the other two methods.

We also observe that the MLP-based method yields poor performance on ImageNet-1k models, given the relatively high overfitting level of these models.
We suspect that the dimensions of the full posterior inputs commonly used in these models are too large for the MLP-based attack.
To accommodate the large input dimension, we select only the top-3 largest posteriors as input to make the attack model more sensitive to important information in the posterior.
Different from observations in previous work~\cite{SZHBFB19}, \autoref{figure:membership_mlp_methods} shows the MLP-based attack improves significantly on ImageNet-1k models and reaches similar performance as the modified entropy attack.
The improvement is less prominent in the CIFAR-100 and CUB-200-2011 models, where the target datasets have a smaller number of classes.
This means that the attack performance of the developed methods may vary significantly when evaluated on more complex datasets.
While it can be resource-intensive to conduct experiments on more complex datasets, researchers can consider using a few trained benchmark models for attack evaluation in the future.

\begin{figure}[!t]
\centering
\includegraphics[width=0.73\columnwidth]{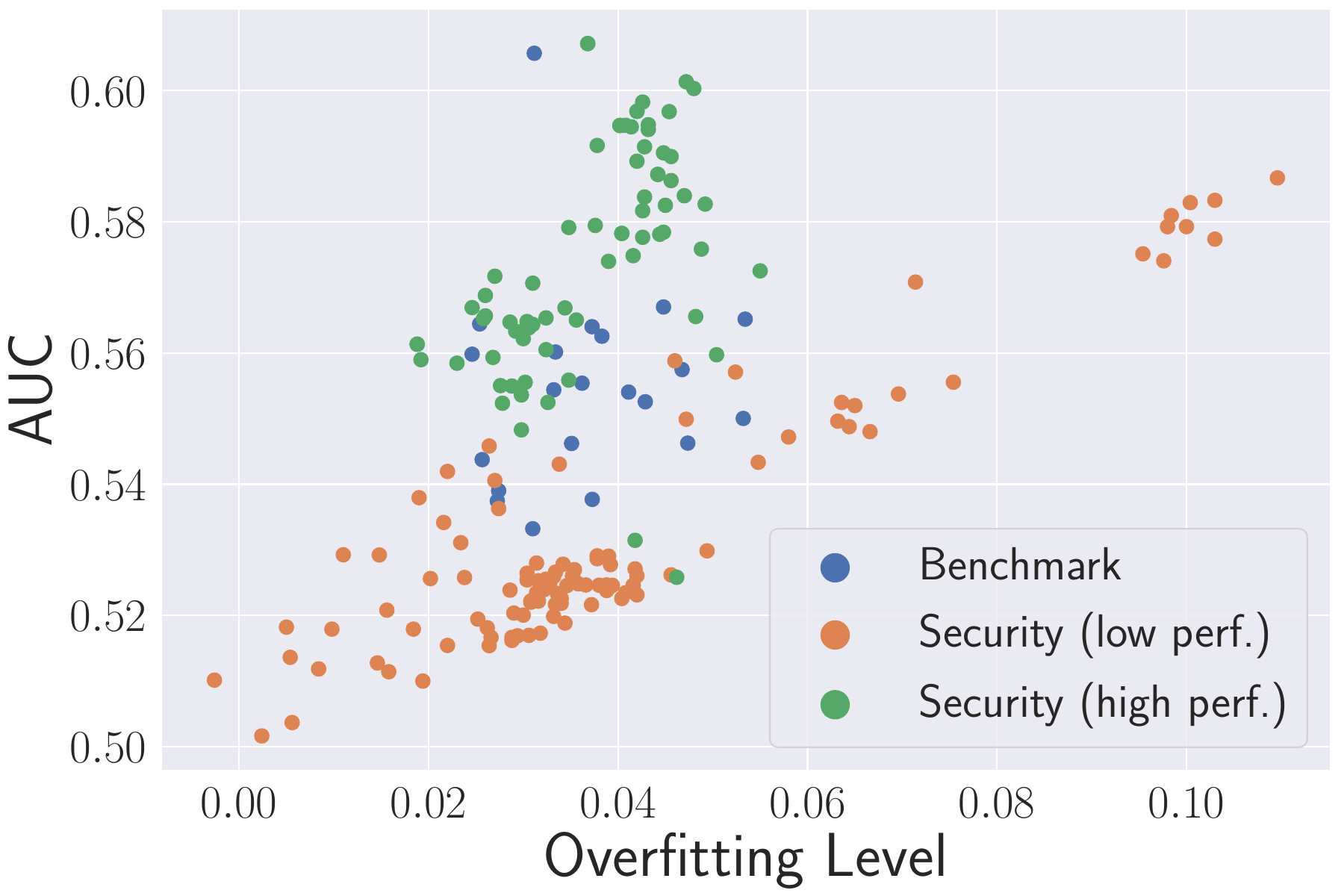}
\caption{The membership inference performance (AUC) with respect to the target model’s overfitting level on CIFAR-10 benchmark and security models.}
\label{figure:meminf_cifaradv_compare}
\end{figure}

\mypara{Benchmark vs.\ Security Models}
Similar to our model stealing analysis, we also compare the membership inference attack performance between benchmark and security models.
\autoref{figure:meminf_cifaradv_compare} shows the performance of the modified entropy attacks.
We first observe that both types of models generally show a similar positive correlation (0.609) with the overfitting level, indicating that the overfitting level, indeed, is the primary indicator of membership vulnerability.
Moreover, we find that the two clusters of security models, which have relatively high and low target task performance, respectively (see \autoref{figure:violin_target_accuracy}), also react differently to membership inference attacks.
The low-performing security models appear less vulnerable to the attack than both the high-performing ones and benchmark models despite having a similar overfitting level.
As a result, the models that are not trained adequately on the target tasks can potentially appear to be less vulnerable and lead to underestimated risks in evaluation.

\begin{figure}[!t]
\centering
\includegraphics[width=0.73\columnwidth]{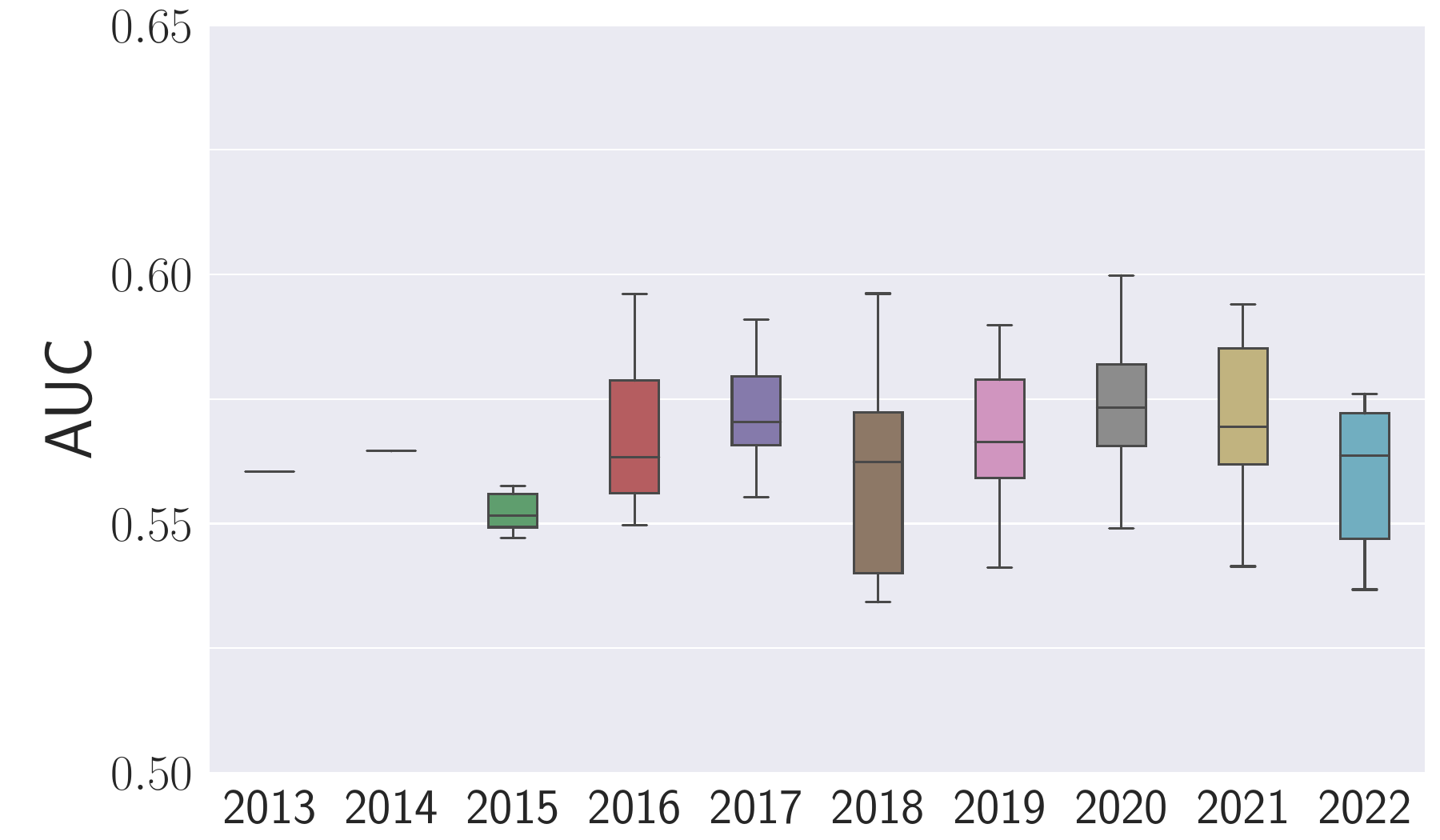}
\caption{The membership inference performance (AUC) with respect to the publication year.}
\label{figure:meminf_year}
\end{figure}

\begin{figure}[!t]
\centering
\includegraphics[width=0.73\columnwidth]{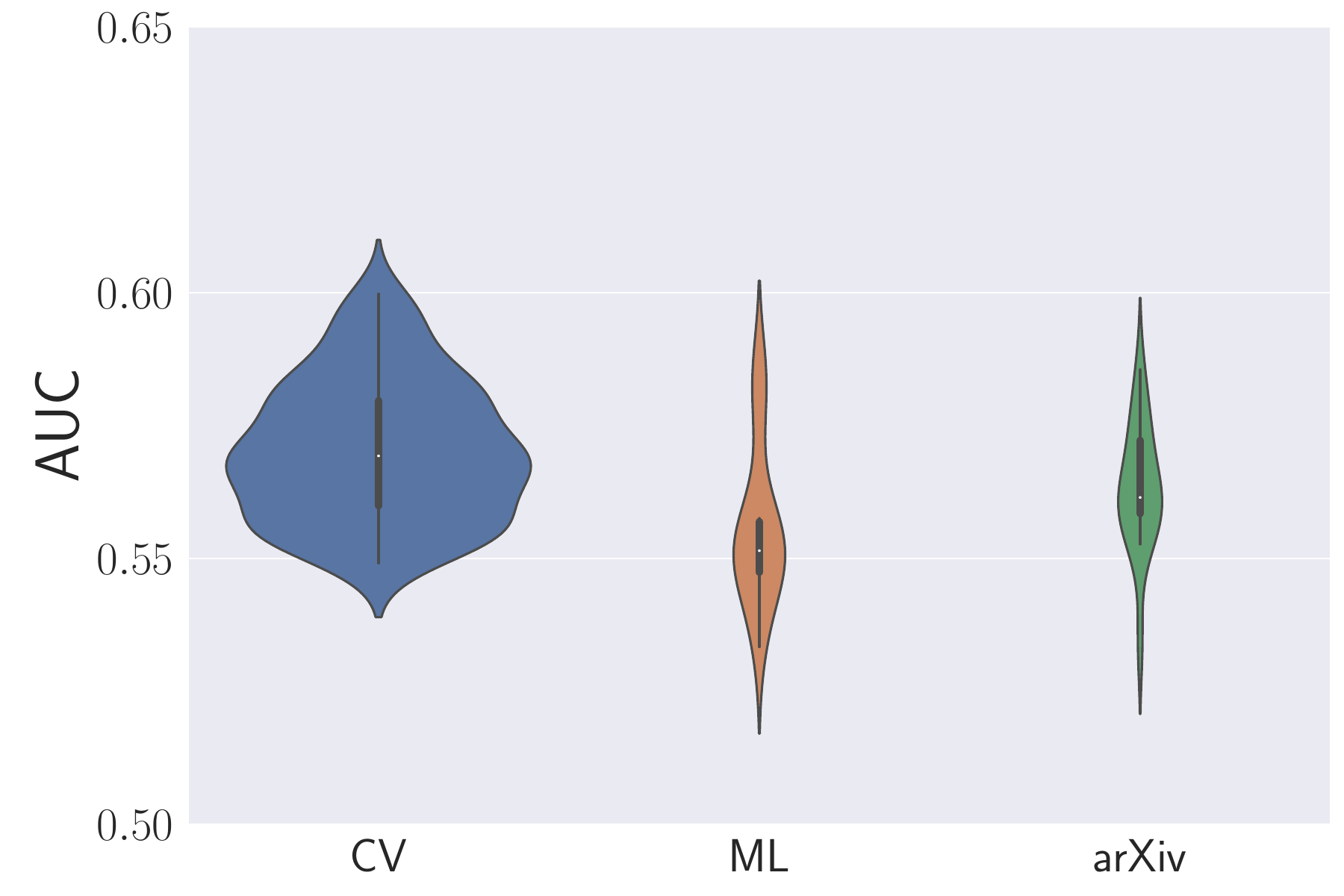}
\caption{The membership inference performance (AUC) with respect to the conference type.}
\label{figure:meminf_conftyp}
\end{figure}

\mypara{Metadata}
For membership inference, we also examine the correlation between the attack performance (modified entropy) and two types of models' metadata, publishing time and conference type, using ImageNet-1k benchmark models.
\autoref{figure:meminf_year} shows that, unlike model stealing attacks, newer models are not more (or less) secure to membership inference attacks compared to older ones.
Meanwhile, in \autoref{figure:meminf_conftyp}, we also do not observe significant performance differences in membership inference on models from different venues.

\subsection{Backdoor Detection}

We next focus on backdoor detection on the public models.
We first emphasize that while it is crucial for the backdoor detection techniques to identify the backdoored model accurately, the techniques' practicality also depends on having an acceptable low false positive rate.
Thus, we examine the false positive rate of three widely-used backdoor detections on models from \database.
Note that since these methods rely on finding easily misclassified labels/images iteratively, the computation cost can be very high.
Therefore, we only conduct evaluations on CIFAR-10 and SVHN models.
Furthermore, since we aim to examine the false positive rate of these techniques, we only consider benchmark models.
The reason is that these benchmark models are (almost) unlikely to contain backdoors.

\begin{table}[!t]
\centering
\tabcolsep 4pt
\scalebox{0.9}{
\begin{tabular}{c|rrr}
\toprule
\textbf{Detection Method} & \textbf{CIFAR-10} & \textbf{SVHN} & \textbf{Runtime}\\
\midrule
Neural Cleanse  & 20.9\% & 13.7\% & 802.1s\\ 
STRIP & 0.0\% & 0.0\% & 32.1s\\
NEO & 0.0\% & 0.0\% & 18.0s\\
\bottomrule
\end{tabular}
}
\caption{Backdoor detection performance (false positive rate) on CIFAR-10 and SVHN models. 
Runtime is from CIFAR-10's ResNet-18 model.}
\label{table:backdoor}
\end{table}

\begin{figure}[!t]
\centering
\begin{subfigure}{0.23\textwidth}
\centering
\includegraphics[width=0.5\textwidth]{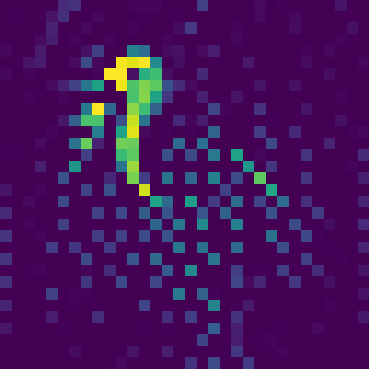}
\caption{Trigger Mask CIFAR-10}
\label{figure:trigger_map_cifar}
\end{subfigure}%
\hfill
\begin{subfigure}{0.23\textwidth}
\centering
\includegraphics[width=0.5\textwidth]{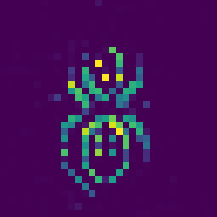}
\caption{Trigger Mask SVHN}
\label{figure:trigger_map_svhn}
\end{subfigure}%
\hfill
\begin{subfigure}{0.23\textwidth}
\centering
\includegraphics[width=0.5\textwidth]{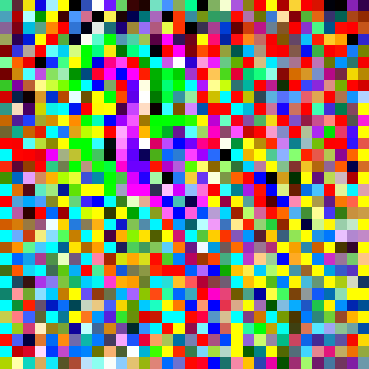}
\caption{Trigger Image CIFAR-10}
\label{figure:trigger_cifar}
\end{subfigure}
\hfill
\begin{subfigure}{0.23\textwidth}
\centering
\includegraphics[width=0.5\textwidth]{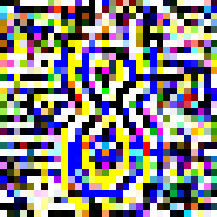}
\caption{Trigger Image SVHN}
\label{figure:trigger_svhn}
\end{subfigure}
\caption{The trigger masks (a and b) and patterns (c and d) generated by the backdoor detection method Neural Cleanse.}
\label{figure:trigger_pattern}
\end{figure}

\mypara{Model Inspection}
For the model inspection method, Neural Cleanse has very high detection rates, specifically 20.9\% for CIFAR-10 models and 13.7\% for SVHN models, shown in \autoref{table:backdoor}.
To determine whether the detection is false positive, we examine the trigger patterns.
The method provides trigger patterns generated through optimization and shows both a trigger image and a mask image of the trigger location.
The mask area is selected as an outlier through the detection process, i.e., much smaller than others to cause misclassification.
\autoref{figure:trigger_pattern} shows two examples detected by Neural Cleanse.
The examples, however, do not resemble any trigger patterns.
More specifically, the mask area is still too large and generally covers the key areas in the current class of images.
For example, the SVHN trigger mask clearly shows the digit 8, which means almost all of the area has to be altered to cause misclassification and is, therefore, not a true trigger.
The CIFAR-10 example similarly shows the outline of a bird (which is the label of the class).
We can confirm that the detected triggers are indeed all false positives.
The false positive rates on these public models greatly exceed the results on their experiment models presented in the original work~\cite{WYSLVZZ19}.
The generated trigger patterns do help users eliminate the false positive samples easily, yet they require manual intervention.

Besides, since Neural Cleanse iteratively optimizes the trigger pattern and evaluates the change in prediction results, the run time on our GPU cluster (an NVIDIA DGX-A100 server) is at least 25 times longer than the other two methods.
The current run-time evaluation actually benefits Neural Cleanse by evaluating a model with a simple architecture (ResNet-18) trained on a small dataset (CIFAR-10).
The detection algorithm's run time will scale with not only the model's computation complexity but also the number of classes in the dataset.
For datasets with more classes, such as ImageNet-1k, the method will become infeasible since the run time will become at least 100 times more than the current setting, even if we assume the time cost for each label's iteration remains the same.
This significant resource requirement can hinder the method's practicality in the real world.

\mypara{Input Filtering}
For the input filtering methods, i.e., STRIP and NEO, we adopt a subset of the test set for evaluation.
Since we choose the images ourselves, we can ensure there is no backdoor, and thus, the method should correctly identify the images as clean.
Our experiments show that the two methods are effective in terms of low false positive rates.
None of the models is detected as having backdoors.
Noticeably, both methods have detection values that are much lower than their respective thresholds, which further indicates the method is effective in avoiding over-detection.
The computation cost is also significantly lower compared to Neural Cleanse and realistically allows real-world deployment.

\subsection{Result Summary}

Thanks to \database, we are able to perform an extensive evaluation for model stealing, membership inference, and backdoor detection on a large set of public models, which, to the best of our knowledge, has not been done before.
Our analyses confirm some results from previous works but on a much larger scale, discover some new insights, and show some of the previous results obtained from researchers' self-trained models can vary on public models.

First of all, we find that the model stealing attack can perform especially poorly on certain datasets, such as CUB-200-2011, in contrast to target models (with the same architecture) trained on other datasets.
Using an out-of-distribution auxiliary dataset also does not improve the attack on our public models.
Furthermore, we demonstrate that the model stealing performance negatively correlates with the model's target task performance and is too low to be effective on some modern high-performing models.
Unlike previous works~\cite{OSF19}, we find using a more complex surrogate model does not improve the attack performance.
These observations imply that the proposed methods, which perform well under experimental conditions, can become inadequate on public models.

As for membership inference, we make a similar observation, as shown in previous works, that the attack performance positively correlates with the victim model's overfitting level.
Additionally, we find methods that perform well on experiment datasets do not guarantee similar performance on more difficult datasets.
In contrast to previous work's~\cite {SZHBFB19} results, the MLP-based attack performs differently on models trained with data that contains a large number of classes (e.g., ImageNet-1k) when using different input methods.

Additionally, for both model stealing and membership inference, we compare the behavior of security models to that of benchmark models.
We notice the security models with low target task performance can react drastically differently to both attacks.
More concretely, model stealing attack positively correlates with the target model's performance, and membership inference is less effective given a similar overfitting level.
The two observations cannot be made on security models with similar target task performance as benchmark models, and thus, we suspect the training level mainly causes the different attack behavior.
We hope to emphasize the necessity of training the target or victim models ``properly,'' i.e., close to the architecture's maximum performance on the target task or using public models as target models for evaluation.

Finally, for backdoor detection, we evaluate the methods' false positive rates on a large number of public models.
This allows us to report our observation on the high false positive rate of Neural Cleanse with more confidence, which may be difficult to conclude from just a few test models.
The resource requirement or run time of the detection method should also be taken into account when developing detection methods.

\section{Related Works}
\label{section:related works}

\mypara{Model Stealing}
Several previous works have shown that machine learning models can be vulnerable to model stealing attacks~\cite{TZJRR16,JCBKP20,CJM20,WG18,OASF18,OSF19,KTPPI20,SHHZ22,SHYBZ23,ZHSWZ23}.
In general, model stealing attacks focus on either extracting the target model's parameters~\cite{CJM20,TZJRR16,JCBKP20} or functionalities~\cite{OSF19,KTPPI20,SHHZ22,JCBKP20,SHYBZ23,ZHSWZ23}.
Tram{\`e}r et al.~\cite{TZJRR16} propose the first model stealing attacks against black-box ML models with prediction API.
Orekondy et al.~\cite{OSF19} develop Knockoff Nets that can steal the functionality of the given target model and leverage a reinforcement learning approach to improve the query sample efficiency.
Model stealing attacks have been applied to different machine learning applications such as BERT-based APIs~\cite{KTPPI20}, Graph Neural Networks (GNNs)~\cite{SHHZ22}, and Contrastive Learning~\cite{SHYBZ23}.

\mypara{Membership Inference}
Existing works on membership inference rely heavily on self-trained models to ensure membership information.
Shokri et al.~\cite{SSSS17} develop the first membership inference attacks against ML models.
Salem et al.~\cite{SZHBFB19} relax such assumptions of~\cite{SSSS17} by using only one shadow model to establish the attack.
Nasr et al.~\cite{NSH19} further investigate the membership leakage via the white-box access to the target model.
Song and Mittal~\cite{SM21} observe that metric-based attacks can have similar or even better performance than previous attacks that leverage ML-based attack models.
Label-only attacks~\cite{LZ21,CTCP21} have been proposed for a more difficult scenario where the adversary can only obtain predicted labels instead of the posteriors from the target model.

\mypara{Backdoor Detection}
Chen et al.~\cite{CLLLS17} propose the first backdoor attack and, more specifically, the first targeted backdoor attack using data poisoning.
Recently, numerous works have introduced detection methods for both targeted and untargeted backdoored models.
Similar to Neural Cleanse~\cite{WYSLVZZ19} examined in this paper, many previous works detect backdoors by inspecting the models~\cite{CFZK19,HAS19,GWXDS19,LDG18,LLTMAZ19}.
Others such as Cohen et al.~\cite{CRK19} detect trigger inputs at inference time like STRIP~\cite{GXWCRN19} and NEO~\cite{UPWLRC22}.

\mypara{Public Model Analysis and Evaluations}
There is not much work on analyzing public models' behaviors, especially from the security and privacy angle.
Gavrikov and Keuper~\cite{GK22} analyze the properties of the distribution of 3x3 convolution filter kernels from hundreds of trained models.
Schürholt et al.~\cite{STKGB22} present a dataset of 50,360 systematically generated neural network models for future model property research.
This collection of trained models focuses more on providing diverse training trajectories through different combinations of hyperparameters, and thus, models do not necessarily reflect the ones publicly available online.

\mypara{Large-Scale Evaluation of ML Security and Privacy}
Another related topic is the measurement study on the security and privacy risks of machine learning models.
Liu et al.~\cite{LWHSZBCFZ22} examine four inference attacks using five model architectures trained on four datasets.
Pang et al.~\cite{PZGXJCW20} develop TrojanZoo, an open-source platform for evaluating backdoor attacks/defenses.
For evasion attacks, a selection of previous works \cite{LJZWWLW19,PFCGFKXSBRMBHZJLSGUGDBHRLM18,CASDFCMH21} propose security analyses and benchmark platforms for generating and defending adversarial examples.
These works, however, aim at developing the toolbox for future risk assessment but do not include evaluation analyses on a large set of public models.

\section{Conclusion}
\label{section:conclusion}

In this paper, we collect and annotate an extensive database of public models, namely \database, for privacy and security research in machine learning.
We examine these public models with model stealing, membership inference, and backdoor detection.
Compared to the results in previous works obtained from researchers' self-trained models, we discover some new insights on ML attacks/defenses with \database.
We will share \database with the community and recommend future researchers include experiments on public models to demonstrate their methods' efficacy.

\medskip
\mypara{Acknowledgments}
We thank all anonymous reviewers for their constructive comments. 
This work is partially funded by the European Health and Digital Executive Agency (HADEA) within the project ``Understanding the individual host response against Hepatitis D Virus to develop a personalized approach for the management of hepatitis D'' (D-Solve) (grant agreement number 101057917).

\bibliographystyle{plain}
\bibliography{normal_generated_py3}

\appendix
\section*{Appendix}

\section{Annotating \database}
\label{section:categories}

To annotate relevant information about the models in \database, we utilize a two-step iterative coding process, which has been widely adopted in various fields such as usable security, social computing, and psychology~\cite{C06, LFH17}.
Concretely, 5 researchers first traverse all models and their relevant information and independently assign initial codes about the categories.
They then work together to discuss the initial codes and their interconnections.
Disagreement is solved through discussion and a majority vote.
After agreeing on a final codebook (see \autoref{table:codebook}), they code the entire database again.
Our results show high inter-coder agreement (Cohen's kappa = 0.915).

\begin{table*}[!t]
\centering
\scalebox{0.9}{
\begin{tabular}{p{0.2\linewidth} | p{0.4\linewidth} |p{0.4\linewidth}}
\toprule
\textbf{Code} & \textbf{Description} & \textbf{Examples}\\
\midrule
Task Type  &  The type of tasks the model designed to perform & classification, detection, segmentation \\
\midrule
Dataset  & The dataset used to train the model & CIFAR-10, ImageNet-1k\\
\midrule
Topic Category & The types of subjects in the given dataset & natural scenery, medical scans, satellite images\\
\midrule
Training Set Size & The number of samples in the training data &  50000, 5994, 1281167\\
\midrule
Number of Classes  &  Number of classes in the dataset & 10, 200, 1000\\
\midrule
Input Image Dimensions & The image size and number of channels & 28$\times$28$\times$1, 32$\times$32$\times$3, 224$\times$224$\times$3\\
\midrule
Publishing Venues &  The conference or journal where the paper that first introduced the model is published (if applicable) & CVPR, ICCV, NeurIPS\\
\midrule
Venue Type &  The type/topic of the conference/journal & computer vision, machine learning, preprint\\
\midrule
Publishing Year & The year of the publication for the paper (models can be released online later) & 2013, 2017, 2020\\
\midrule
Number of Authors & The number of authors appeared on the paper that published the given model architecture & 2, 3, 8\\
\midrule
Affiliation & The author's affiliation  & Stanford University, Google, Max Planck Institute\\
\midrule
Affiliation Type & The types of affiliation & Companies, Research Institutions, Universities\\
\midrule
Country & The country of where the authors' affiliations are established &  USA, Germany, China, UK\\
\bottomrule
\end{tabular}
}
\caption{The final codebook for \database annotation.}
\label{table:codebook}
\end{table*}

\section{Model Stealing Query Budget}
\label{section:steal_budget}

\begin{figure}[!t]
\centering
\includegraphics[width=0.73\columnwidth]{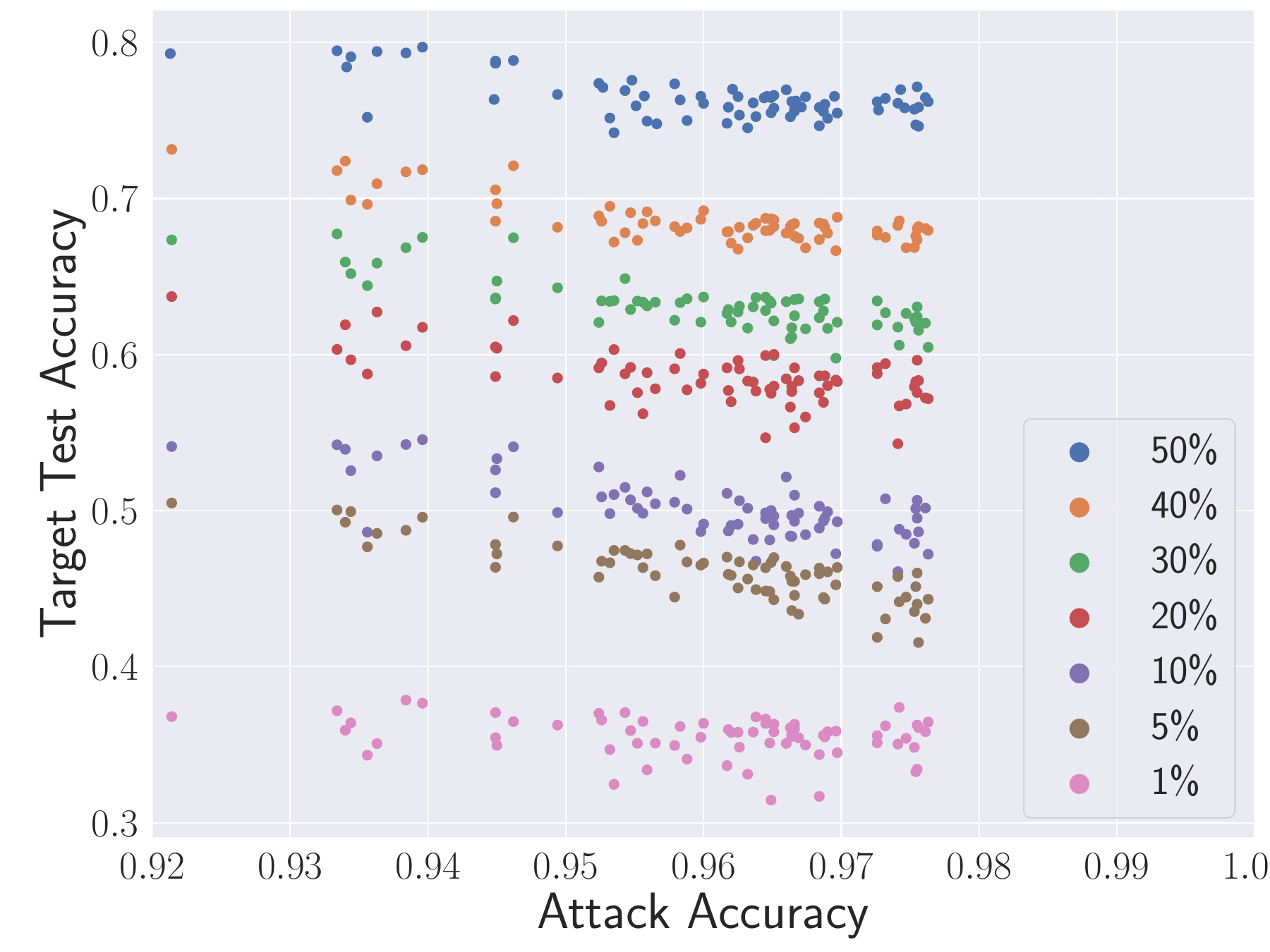}  
\caption{The relationship between the target task accuracy and the attack accuracy for model stealing on CIFAR-10 models when using different query budgets.}
\label{figure:steal_budget}
\end{figure}

When using the partial training set to steal target models, we assume the attacker has 50\% of the original training data.
We also evaluate how different query budget affects the attack performance on CIFAR-10 models.
Similar to previous works~\cite{OSF19, TZJRR16}, we find positive correlations between the size of training data and attack performance, as shown in \autoref{figure:steal_budget}.

\section{Evasion Attacks}
\label{section:evasion_attacks}

Although evasion attacks is not the focus of this paper, we still include some baseline results for the benefit of future researchers in this domain.
We examine the standard FGSM~\cite{GSS15} attack in white-box and black-box settings.
In the white-box setting, the adversary has full access to the model and generates adversarial noise directly.
While in the black-box setting, the adversary cannot obtain the gradient from the target model directly but uses an auxiliary model (ResNet-18 trained on the same data) to generate the noise.

\begin{figure*}[!t]
\centering
\begin{subfigure}{0.33\textwidth}
\centering
\includegraphics[width=\textwidth]{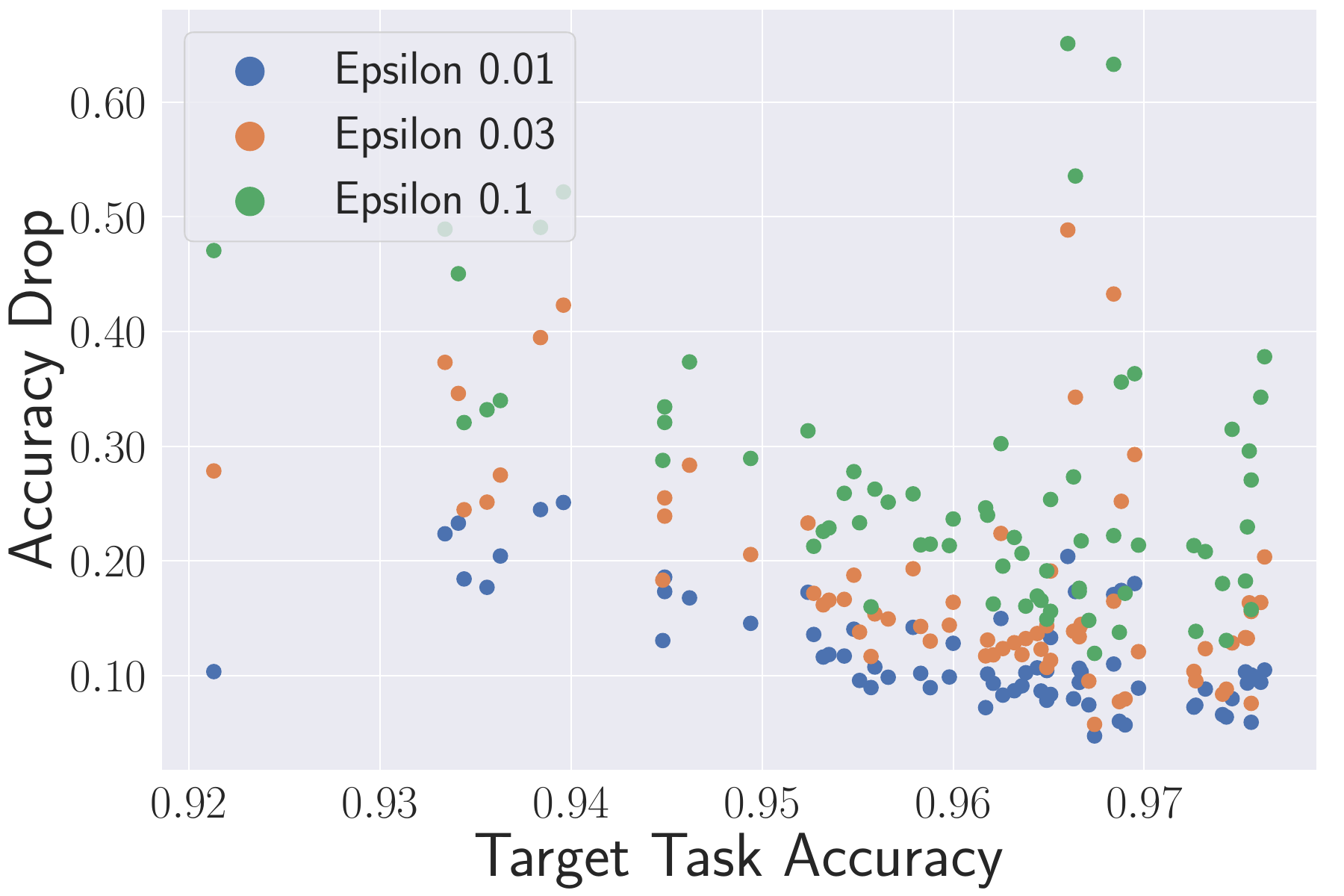}
\caption{CIFAR-10}
\label{figure:fgsm_cifar10}
\end{subfigure}%
\hfill
\begin{subfigure}{0.33\textwidth}
\centering
\includegraphics[width=\textwidth]{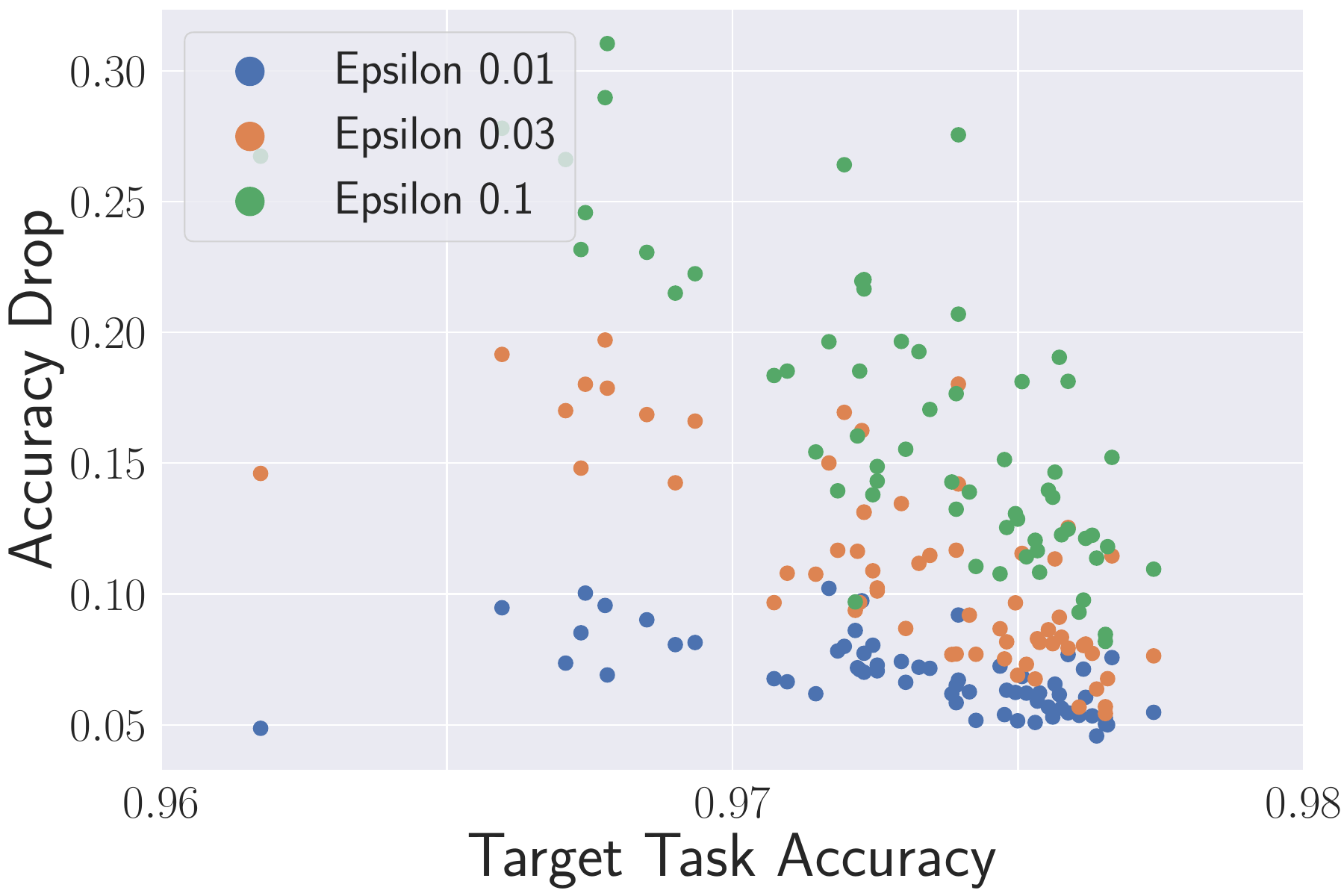}
\caption{SVHN}
\label{figure:fgsm_svhn}
\end{subfigure}%
\hfill
\begin{subfigure}{0.33\textwidth}
\centering
\includegraphics[width=\textwidth]{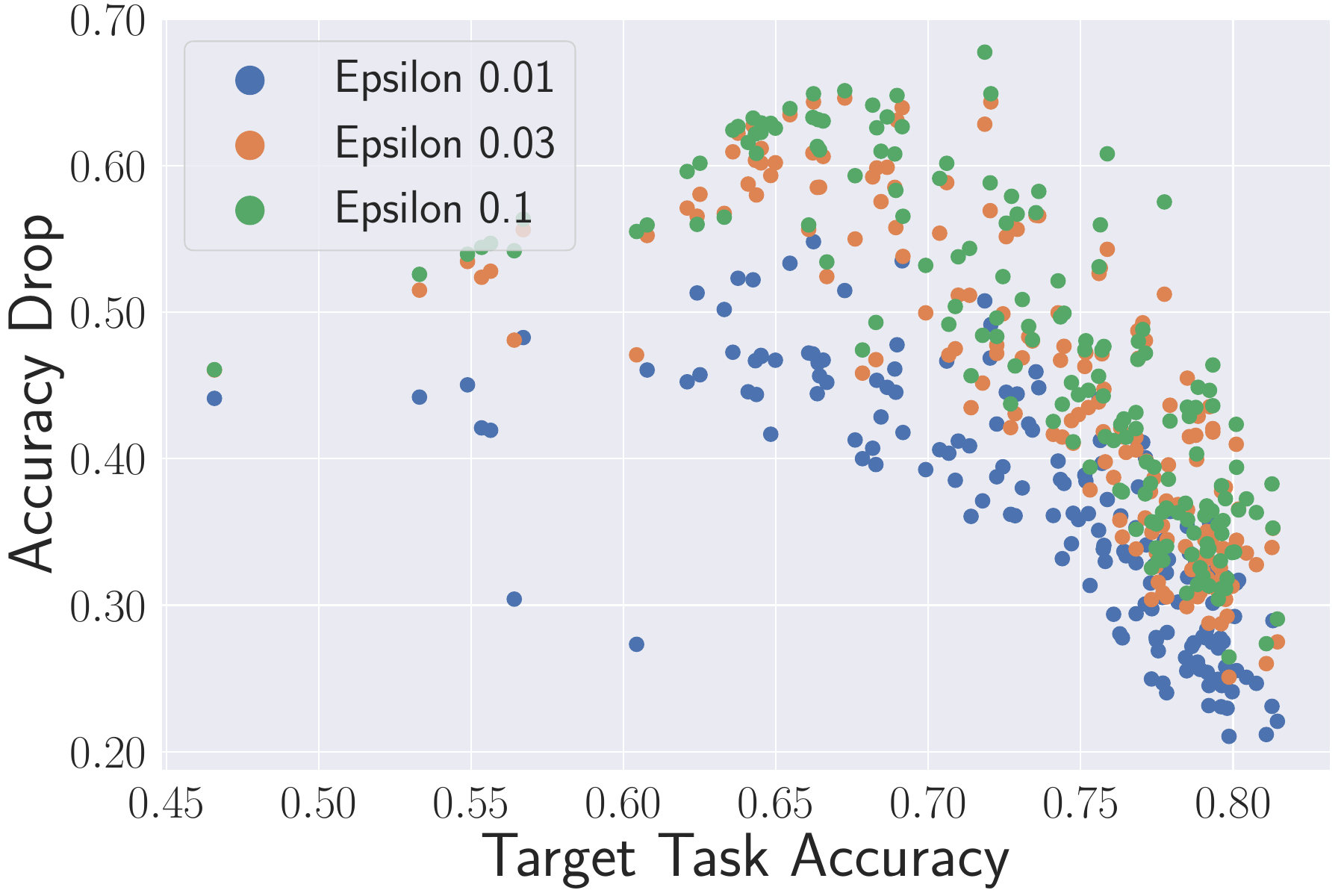}
\caption{ImageNet-1k}
\label{figure:fgsm_large}
\end{subfigure}%
\caption{The relationship between the evasion attack effectiveness (target task accuracy drop) and the target model's task accuracy across various benchmark models under white-box setting with different epsilons.}
\label{figure:FGSM}
\end{figure*}

\begin{figure*}[!t]
\centering
\begin{subfigure}{0.33\textwidth}
\centering
\includegraphics[width=\textwidth]{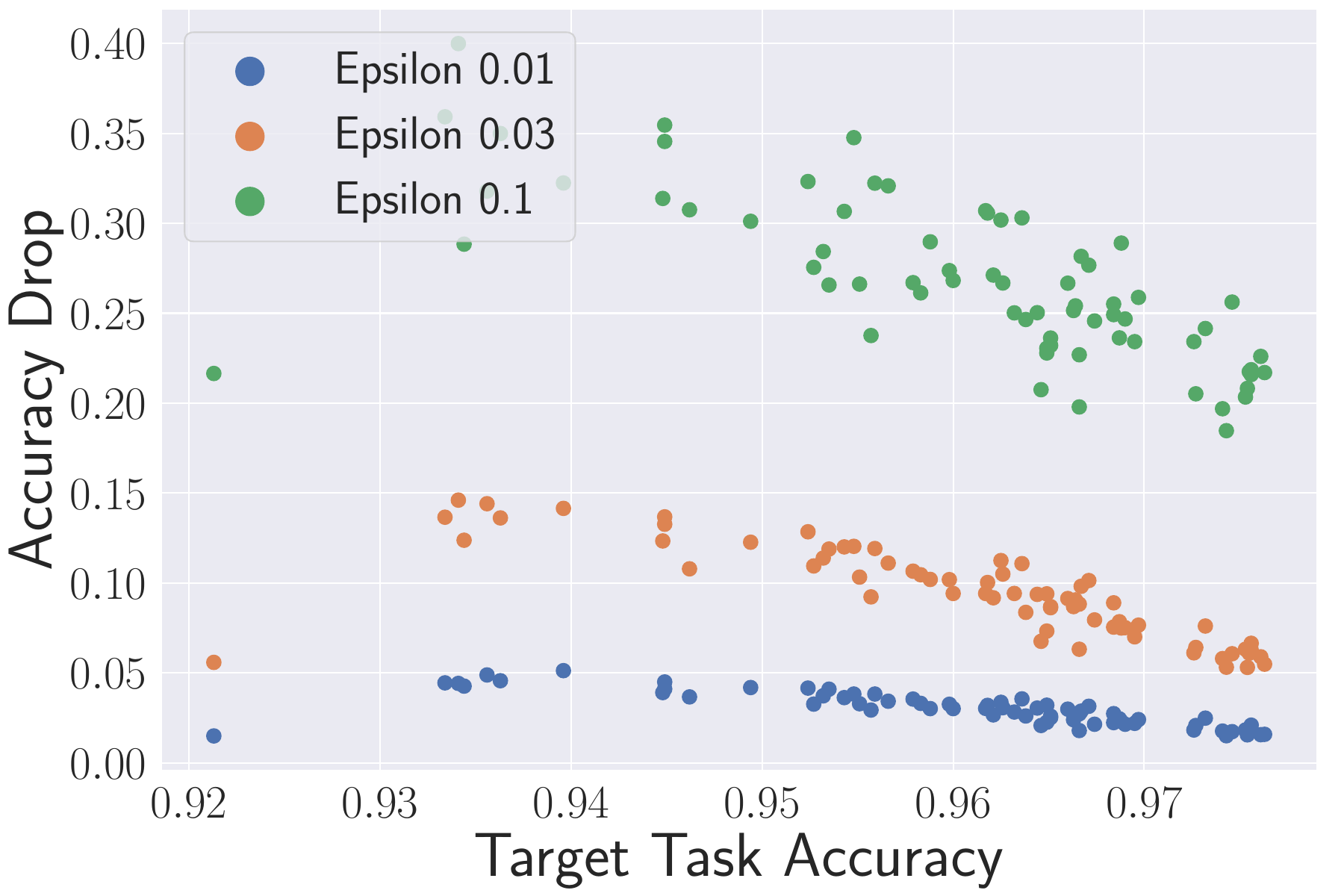}
\caption{CIFAR-10}
\label{figure:fgsm_transfer_cifar10}
\end{subfigure}%
\hfill
\begin{subfigure}{0.33\textwidth}
\centering
\includegraphics[width=\textwidth]{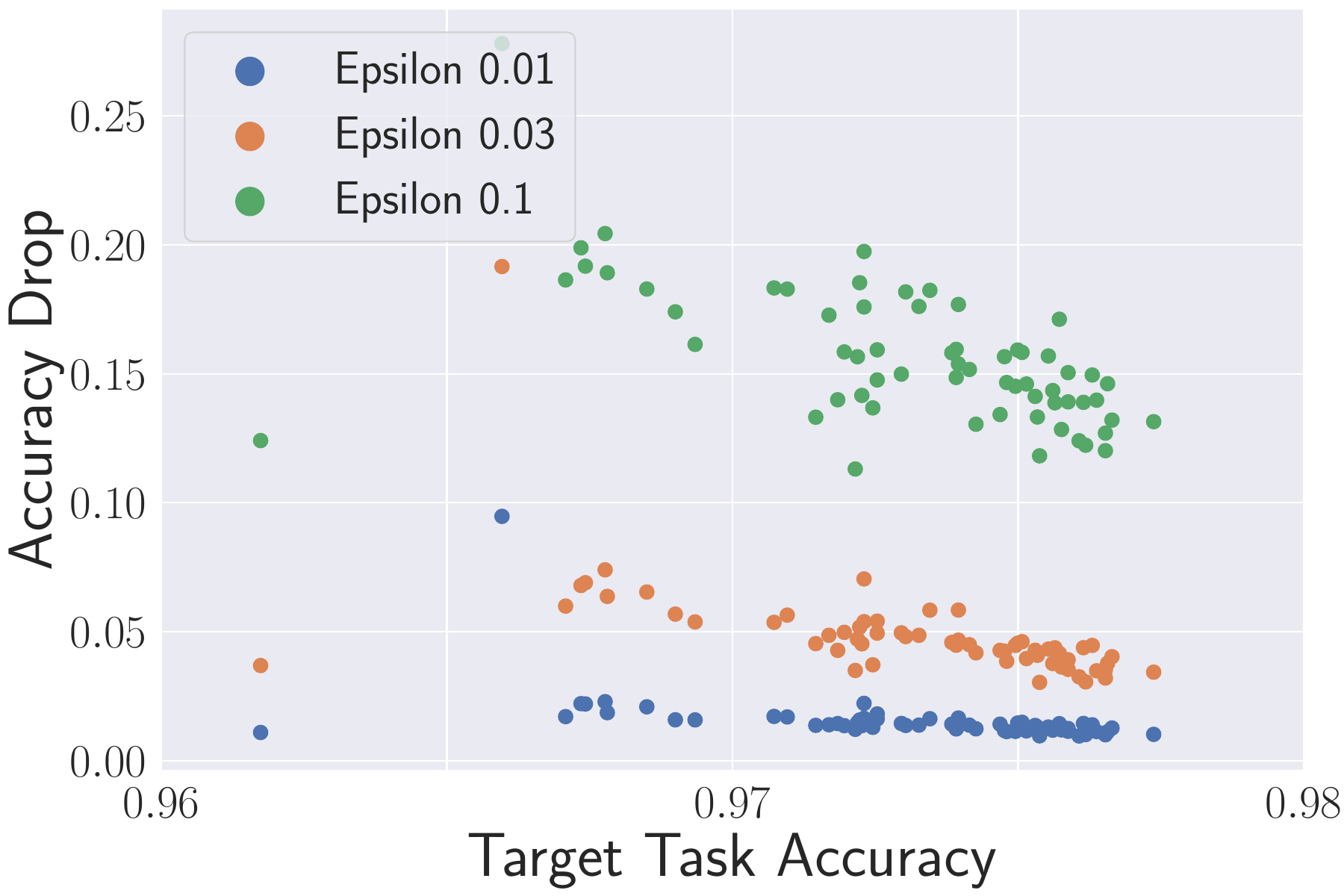}
\caption{SVHN}
\label{figure:fgsm_transfer_svhn}
\end{subfigure}%
\hfill
\begin{subfigure}{0.33\textwidth}
\centering
\includegraphics[width=\textwidth]{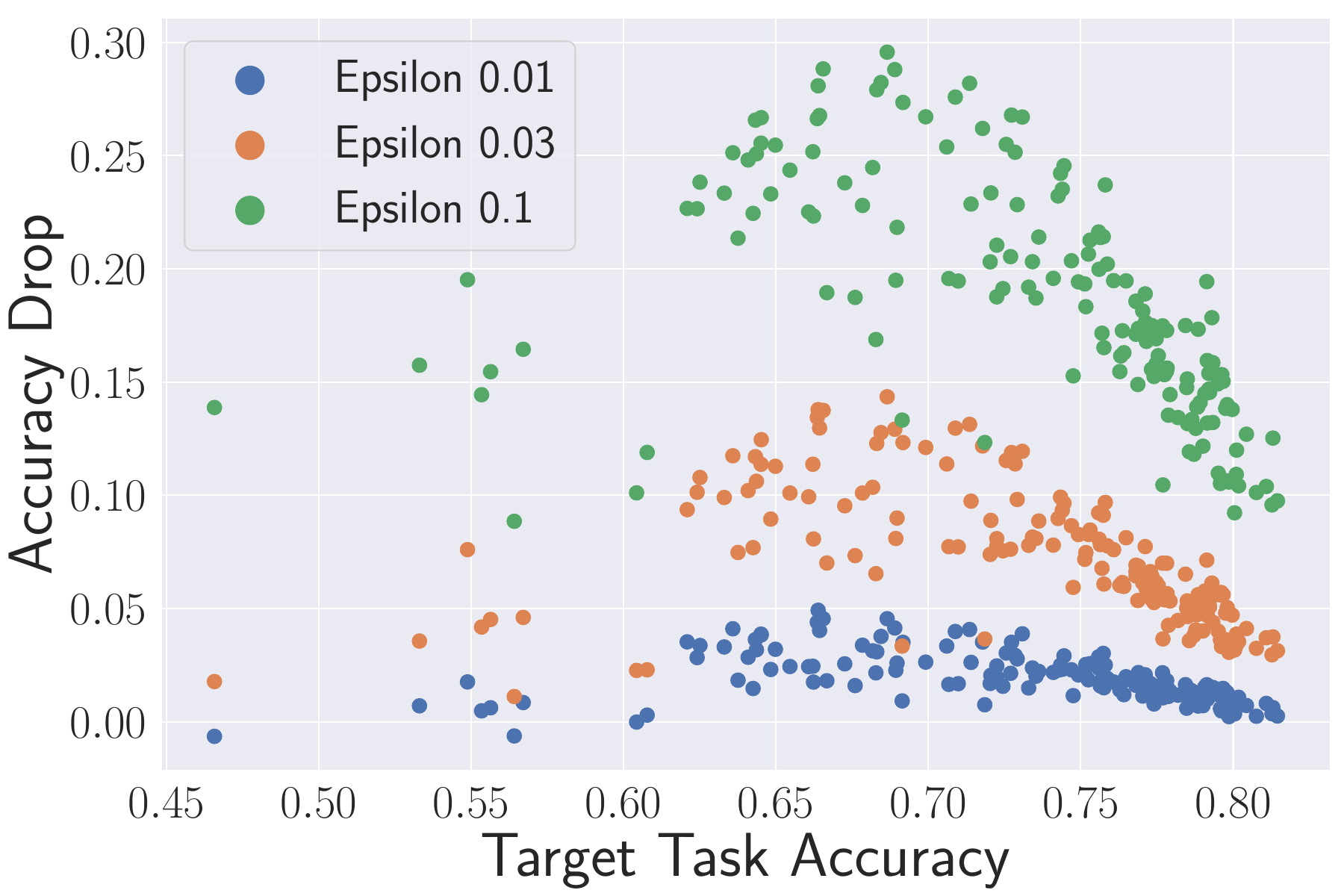}
\caption{ImageNet-1k}
\label{figure:fgsm transfer_imgnet}
\end{subfigure}%
\caption{The relationship between the evasion attack effectiveness and the target model's task accuracy across various benchmark models under the black-box setting with different epsilons.}

\label{figure:FGSM_transfer}
\end{figure*}

We present the accuracy drop on the target task in three different noise levels with epsilon at 0.01, 0.03, and 0.1 in \autoref{figure:FGSM}.
We find there is a negative correlation between the attack effectiveness and the model's original target accuracy (e.g., the Pearson correlation coefficient is -0.491 on ImageNet-1k models).
The method becomes ineffective on large models, even in the white-box setting with relatively large noise.
For instance, the CIFAR-10 ResNet-1202 model only suffers an accuracy drop of 0.161 in the white-box setting with epsilon 0.1, compared to the much larger 0.491 that of ResNet-20 from the same model family.
Under the black-box setting, shown in \autoref{figure:FGSM_transfer}, the effectiveness is very limited, but the negative correlation is still present.
At a low noise level (epsilon 0.01), the accuracy drop of the attack is generally lower than 0.05.
The negative correlation implies that the method that works on smaller models does not necessarily work as effectively on larger ones.
Future researchers should consider selecting a few public models in \database that are particularly ineffective with baseline attacks to demonstrate the effectiveness and improvement of the proposed method.

\section{Limitation}
\label{section:limitation}

Our project is not without some limitations.
When searching for models from academic papers, we focus on papers from top-tier conferences and thus potentially omit some models.
We believe models from top conferences are more likely to be adopted by the community.
Our analysis is also conducted with PyTorch models only.
In later updates, we hope to include models from other platforms, such as TensorFlow.
Given the large number of models, we still believe the result is representative even with these biases.

Using pre-trained models can limit the evaluation of some attacks against ML models.
To ensure thorough analysis, membership inference attacks usually require control over the training process.
However, we advocate for including a few case studies with public models to show that the proposed methods indeed perform similarly on larger models.

\end{document}